\documentclass[showpacs, twocolumn, prb]{revtex4}
\usepackage{amssymb}

\usepackage{graphicx}
\usepackage{color}

\input{tcilatex}

\begin{document}

\title{Effective anisotropies and energy barriers of magnetic nanoparticles with N\'eel surface anisotropy}
\author{R. Yanes and O. Chubykalo-Fesenko}
\affiliation{Instituto de Ciencia de Materiales de Madrid, CSIC,
Cantoblanco, 28049 Madrid, Spain}
\author{H. Kachkachi}
\affiliation{ \mbox{Groupe d'Etude de la Mati\`ere Condens\'ee,
Universit\'e de Versailles St. Quentin,}\\ \mbox{CNRS UMR8635, 45
av. des Etats-Unis, 78035 Versailles, France}\\}
\author{D. A. Garanin}
\affiliation{\mbox{Department of Physics and Astronomy, Lehman
College, City University of New York,} \\ \mbox{250 Bedford Park
Boulevard West, Bronx, New York 10468-1589, U.S.A.}
 }
\author{R. Evans and R. W. Chantrell}
\affiliation{Department of Physics, University of York,
Heslington, York YO10 5DD, UK}

\begin{abstract}
Magnetic nanoparticles with N\'{e}el surface anisotropy, different
internal structures, surface arrangements and elongation are
modelled as many-spin systems. The results suggest that the energy
of many-spin nanoparticles cut from cubic lattices can be
represented by an effective one-spin potential containing uniaxial
and cubic anisotropies. It is shown that the values and signs of
the corresponding constants depend strongly on the particle's
surface arrangement, internal structure and elongation. Particles
cut from a simple cubic lattice have the opposite sign of the
effective cubic term, as compared to particles cut from the
face-centered cubic lattice. Furthermore, other remarkable
phenomena are observed in nanoparticles with relatively strong
surface effects: (i) In elongated particles surface effects can
change the sign of the uniaxial anisotropy. (ii) In symmetric
particles (spherical and truncated octahedral) with cubic core
anisotropy surface effects can  change its sign. We also show that
the competition between the core and surface anisotropies leads to
a new energy that contributes to both the $2^\mathrm{nd}-$ and
$4^\mathrm{th}-$order effective anisotropies.

We evaluate energy barriers $\Delta E$ as functions of the
strength of the surface anisotropy and the particle size. The
results are analyzed with the help of the effective one-spin
potential, which allows us to assess the consistency of the widely
used formula $\Delta E/V=\mathcal{K}_{\infty
}+6\mathcal{K}_{s}/D$, where $\mathcal{K}_{\infty }$ is the core
anisotropy constant, $\mathcal{K}_{s}$ is a phenomenological
constant related to surface anisotropy, and $D$ is the particle's
diameter. We show that the energy barriers are consistent with
this formula only for elongated particles  for which the surface
contribution to the effective uniaxial anisotropy scales with the
surface and is linear in the constant of the N\'{e}el surface
anisotropy.
\end{abstract}
\maketitle

\affiliation{Instituto de Ciencia de Materiales de Madrid, CSMC, Cantoblanco, 28049 Madrid, Spain}

\affiliation{ \mbox{Groupe d'Etude de la Mati\`ere Condens\'ee, Universit\'e de Versailles St. Quentin,}\\ \mbox{CNRS UMR8634, 45 av. des Etats-Unis, 78035 Versailles, France}\\}

\affiliation{\mbox{Physics Department, Lehman College, City University of New York,} \\ \mbox{250 Bedford Park Boulevard West, Bronx, New York 10468-1589, U.S.A.}
 }

\affiliation{Department of Physics, University of York, Heslington, York YO10 5DD, UK}

\section{\label{sec:intro}Introduction}

Understanding the thermally-activated switching of magnetic nanoparticles is
crucial to many technological applications and is challenging from the point
of view of basic research. Surface effects have a strong bearing on the
behavior of magnetic nanoparticles. They exist almost inevitably due to many
physical and chemical effects including crystallographic arrangement on the
surface, oxidation, broken surface bonds, existence of surfactants, etc.
Magnetic particles are often embedded in non-magnetic matrices which alter
magnetic properties of the surface in many ways such as additional surface
tension and possible charge transfer.

Consequently, the properties of magnetic particles are not the same as those
of the bulk material, and many experiments have shown an increase in the
effective magnetic anisotropy due to surface effects \cite{jametetal01prl,
jametetal04prb, bokderetal94prl, luisetal02prb}. Quantum \emph{ab-initio}
studies have revealed different anisotropy and magnetic moment at the
surface of magnetic clusters embedded in matrices \cite{xiebla02prb}.
Synchrotron radiation studies have confirmed that both spin and orbital
moments at the surface differ significantly from their bulk counterparts
\cite{binnsetal02jpcm}. Recent $\mu $-SQUID experiments on isolated clusters
\cite{jametetal01prl, jametetal04prb} produced more reliable estimations of
surface anisotropy.

Thermal measurements have become an important part of the characterization
of systems of magnetic nanoparticles. Often, these measurements include a
complex influence of inter-particle interactions. However, in other cases,
measurements on dilute systems can provide information on individual
particles. The results show that even in these cases the extracted
information is not always consistent with the approximation picturing the
particle as a macroscopic magnetic moment, and this is usually attributed to
surface effects.

Experimentally, the enhancement of the anisotropy at the surface often leads
to an increase in the blocking temperature of single-domain particles from
which the values of the energy barriers $\Delta E$ may be extracted. The
influence of the surface manifests itself in the fact that the values $%
\Delta E/V$ are different from that of the bulk, i.e., there is an effective
anisotropy $\mathcal{K}_{\mathrm{eff}}$ that is not exactly proportional to
the particle's volume $V$.

One can expect that the effect of the surface reduces when the
particle size increases. In Ref.~\onlinecite{bokderetal94prl} it
was suggested that the effective anisotropy scales as the inverse
of the particle's diameter $D$ according to
\begin{equation}
\Delta E/V=\mathcal{K}_{\mathrm{eff}}=\mathcal{K}_{\infty }+\frac{6\mathcal{K%
}_{s}}{D}.  \label{K_eff}
\end{equation}
Here $\mathcal{K}_{\infty }$ is the anisotropy constant for
$D\rightarrow \infty $, presumably equal
to the bulk anisotropy, and $%
\mathcal{K}_{s}$ is the ``effective'' surface anisotropy. We would
like to emphasize that this formula has been introduced in an
\emph{ad hoc} manner, and it is far from evident that the surface
should contribute into an effective uniaxial anisotropy (related
to the barrier $\Delta E$) in a simple additive manner.
Actually one cannot expect $\mathcal{K}%
_{s}$ to coincide with the atomistic single-site anisotropy,
especially when strong deviations from non-collinearities leading
to ``hedgehog-like''
structures appear \cite{kacdim02prb}. The effective anisotropy $\mathcal{K}_{%
\mathrm{eff}}$ appears in the literature in relation to the measurements of
energy barriers, extracted from the magnetic viscosity or dynamic
susceptibility measurements. The surface anisotropy should affect both the
minima and the saddle point of the energy landscape in this case. The origin
of $\mathcal{K}_{\mathrm{eff}}$ may be expected to be different from that
obtained from, e. g., magnetic resonance measurements. In the latter case
the magnetization dynamics depends on the stiffness of the energy minima
modified by the surface effects. Despite its \emph{ad hoc} character, Eq. (%
\ref{K_eff}) has become the basis of many experimental studies with the aim
to extract the surface anisotropy from thermal magnetization measurements
(see, e.g., Refs.~\onlinecite{luisetal02prb, tronc03jmmm,
respaudetal98prb}) because of its mere simplicity. Up to now there were no
attempts to assess the validity of Eq. (\ref{K_eff}) starting from an atomistic
surface anisotropy models such as the N\'{e}el surface anisotropy.

The aim of the present paper is to understand the influence of the
N\'{e}el surface anisotropy on the behavior of magnetic particles
with different surface arrangements, shapes and sizes. Although
various crystal structures have been investigated, the overall
particle properties were kept close to Co. With the help of
numerical modelling of magnetic particles as many-spin systems, we
show that the energy of the many-spin particle can be effectively
represented by that of an effective one-spin particle with both
uniaxial and cubic anisotropy terms. The effective anisotropy
constants depend on the surface arrangement of the particle,
crystal structure, and elongation. We numerically evaluate the
energy barriers of many-spin particles and show that they can also
be understood in terms of the effective one-spin approximation.
Incidentally, this allows us to establish the conditions for the
validity of Eq. (\ref{K_eff}) which turns out to correctly
describe the magnetic behavior of elongated particles only.

\section{From the atomistic to the effective energy}

\label{Sec-Atomistic-to-effective}

\subsection{The atomistic model}

\label{sec:basics}

We consider the atomistic model of a magnetic nanoparticle
consisting of $\mathcal{N}$ classical spins $\mathbf{s}_{i}$ (with
$|\mathbf{s}_{i}|=1$) taking account of its lattice structure,
shape, and size \cite{dimwys94prb, kacetal00epjb,
kacgar01physa300, kacgar01epjb, igllab01prb, kacdim02prb,
garkac03prl, kacmah04jmmm, kacgar05springer, kacbon06prb}. The
magnetic properties of the particle can be described by the
anisotropic Heisenberg model
\begin{equation}
\mathcal{H}=-\frac{1}{2}J\sum_{ij}\mathbf{s}_{i}\cdot \mathbf{s}_{j}+\mathcal{H}_{\mathrm{anis}},  \label{eq:Hamiltonian}
\end{equation}
where $J$ is the nearest-neighbor exchange interaction and $\mathcal{H}_{\mathrm{anis}}$ contains core and surface anisotropies.

The surface anisotropy is often thought to favor the spin orientation normal to the surface. This is the so-called transverse anisotropy model
(TSA). However, a more solid basis for understanding surface effects is provided by the N\'{e}el surface anisotropy (NSA), \cite{Neel, vicmac93prbrc, jametetal01prl, garkac03prl} which takes into account the symmetry of local crystal environment at the surface. The simplest expression for the NSA that will be used below is, as the exchange energy, a double lattice sum over nearest neighbors $i$ and $j,$
\begin{equation}
\mathcal{H}_{\mathrm{anis}}^{\mathrm{NSA}}=\frac{K_{s}}{2}\sum_{ij}\left(
\mathbf{s}_{i}\cdot \mathbf{u}_{ij}\right) ^{2},  \label{eq:NSA}
\end{equation}
where $\mathbf{u}_{ij}$ are unit vectors connecting neigboring sites. One can see that for perfect lattices the contributions of the bulk spins to $\mathcal{H}_{\mathrm{anis}}^{\mathrm{NSA}}$ yield an irrelevant constant, and the anisotropy arising for surface spins only because of the local symmetry breaking is, in general, biaxial. For the simple cubic (sc) lattice and the surface parallel to any of crystallographic planes, an effective transverse surface anisotropy arises for $K_{s}>0.$ In all other cases NSA cannot be reduced to TSA. In fact, Eq. (\ref{eq:NSA}) can be generalized so that it describes both surface and bulk anisotropy. It is sufficient to use different constants $K_{s}$ for different bond directions $\mathbf{u}_{ij}$ to obtain a second-order volume anisotropy as well. However, to obtain a cubic volume anisotropy that is fourth order in spin components, a more serious modification of Eq. (\ref{eq:NSA}) is required. Thus, for simplicity, we will simply use Eq. (\ref{eq:NSA}) to describe the surface anisotropy and add different kinds of anisotropy in the core.

For the core spins, i.e., those spins with full coordination, the anisotropy energy $\mathcal{H}_{\mathrm{anis}}$ is taken either as uniaxial with easy axis along $z$ and a constant $K_{c}$ (per atom), that is
\begin{equation}
\mathcal{H}_{\mathrm{anis}}^{\mathrm{uni}}=-K_{c}\sum_{i}s_{i,z}^{2},
\label{eq:CoreAnisotropy}
\end{equation}
or cubic,
\begin{equation}
\mathcal{H}_{\mathrm{anis}}^{\mathrm{cub}}=\frac{1}{2}K_{c}\sum_{i}\left(s_{i,x}^{4}+s_{i,y}^{4}+s_{i,z}^{4}\right) .  \label{CoreAnisCubic}
\end{equation}

Dipolar interactions are known to produce an additional ``shape''
anisotropy. However, in the atomistic description, their role in
describing the spin non-collinearities is negligible as compared
to that of all other contributions. In order to compare particles
with the same strength of anisotropy in the core, we assume that
the shape anisotropy is included in the core uniaxial anisotropy
contribution. We also assume that in the ellipsoidal particles the
magnetocrystalline easy axis is parallel to the elongation
direction.

In magnetic nanoparticles or nanoclusters, the number of surface
spins $N_{s}$ is comparable to or even larger than the number of
core spins $N_{c}.$ In addition, the surface anisotropy has a much
greater strength than the core anisotropy because of the local
symmetry breaking. We use the core anisotropy value typical for
Cobalt, $K_{c}\simeq 3.2\times 10^{-24}$ Joule/atom. Numerous
experimental results \cite{luisetal02prb, respaudetal98prb,
tronc03jmmm, weller95prl} show that the value of the surface
anisotropy in Co particles embedded in different matrices such as
alumina, Ag or Au, as well as in thin films and multilayers, could
vary from $K_{s}\simeq 10^{-4}J$ to $K_{s}\simeq J.$ In the
present study we will consider the surface anisotropy constant
$K_{s}$ as a variable parameter. For illustration we will choose
values of $K_{s}$ 10-50 times larger than that of the core
anisotropy, in accordance with those reported in
Ref.~\onlinecite{luisetal02prb} for Co particles embedded into
alumina matrices. Such values of $K_{s}$ are also compatible with
those reported in Refs.~\onlinecite{skocoe99iop,
urquhartetal88jap, perrai05springer}. All physical constants will
be measured with respect to the exchange coupling $J$, so we
define the following reduced anisotropy constants
\begin{equation}
k_{c}\equiv K_{c}/J,\;\;k_{s}\equiv K_{s}/J. \label{def}
\end{equation}
The core anisotropy constant will be taken as $k_{c}\simeq 0.01$
and $k_{c}\simeq 0.0025$. On the other hand, the surface
anisotropy constant $k_{s}$ will be varied. Note that we are using
atomistic temperature-independent constants for anisotropies that
should not be confused with temperature-dependent micromagnetic
anisotropy constants $K(T). $ The relation between atomistic and
micromagnetic uniaxial and cubic anisotropy constants within the
mean-field approximation can be found in Ref.
\onlinecite{yostac57ptp}.
\subsection{The effective energy}
\label{sec:EffectiveEnergy}
Investigation of the magnetization switching of a particle
consisting of many atomic spins is challenging because of the
multidimensionality of the underlying potential that can have a
lot of minima of different topologies, connected by sophisticated
paths. Examples are the ``hedgehog'' structures realized in the case
of a strong enough surface anisotropy that causes strong
noncollinearity of the spins. Obviously in this case Eq.
(\ref{K_eff}), relevant in the one-spin description of the
problem, cannot be a good approximation. Thus an important
question that arises here is whether it is possible to map the
behavior of a many-spin particle onto that of a simpler model
system such as one effective magnetic moment. Such an analysis is
unavoidable since it is a crucial step in calculating relaxation
rates and thereby in the study of the magnetization stability
against thermally-activated reversal.

In the practically important case of dominating exchange
interaction, and thus only a small noncollinearity of the spins,
the problem dramatically simplifies, so that the initial many-spin
problem can be reduced to an effective one-spin problem (EOSP). In
the first approximation, one can consider spins as collinear and
calculate the contribution of the surface anisotropy to the energy
of the system depending on the orientation of its global
magnetization $\mathbf{m},$ ($\left| \mathbf{m}\right| =1$). The
resulting energy scales as the number of surface spins, is linear
in $K_{s}$ and depends on the crystal structure and shape. This
we refer to as the first-order surface-anisotropy energy
$\mathcal{E}_{1}.$ Together with the core anisotropy energy (per
spin)
\begin{equation}
\mathcal{E}_{c}=\frac{N_{c}}{\mathcal{N}}K_{c}\left\{
\begin{array}{cc}
-m_{z}^{2}, & \text{uniaxial} \\
\frac{1}{2}\left( m_{x}^{4}+m_{y}^{4}+m_{z}^{4}\right) , & \text{cubic}
\end{array}
\right.  \label{EcDef}
\end{equation}
$\mathcal{E}_{1}$ can lead to Eq. (\ref{K_eff}). However, for
crystal shapes such as spheres or cubes $\mathcal{E}_{1}$ vanishes
by symmetry. In Ref. \cite{garkac03prl} it was shown that for an
ellipsoid of revolution with axes $a$ and $b=a(1+\epsilon )$,
$\epsilon \ll 1,$ cut out of an sc lattice so that the ellipsoid's
axes are parallel to the crystallographic directions, the
first-order anisotropy is given by
\begin{equation}
\mathcal{E}_{1}=-K_{\mathrm{ua}}m_{z}^{2},\qquad K_{\mathrm{ua}}\sim -\frac{N_{s}}{\mathcal{N}}K_{s}\epsilon .
\label{eq:EnElong}
\end{equation}
where the $z$ axis is assumed to be parallel to the
crystallographic axis $b$. That is, $\mathcal{E}_{1}$ scales with
particle size as $\sim1/\mathcal{N}^{1/3}\sim 1/D. $ One can see
that, for the uniaxial core anisotropy along the elongation
direction of the ellipsoid, $K_s<0$ and $\epsilon>0,$ Eq.
(\ref{K_eff}) follows. On the contrary, in other cases, as for
example, $\epsilon >0, K_s>0$ Eq. (\ref{K_eff}) does not apply.

If one takes into account the noncollinearity of the spins that
results from the competition of the exchange interaction and
surface anisotropy and is described by the angles of order $\delta
\psi \sim \mathcal{N}^{1/3}K_{s}/J,$ a contribution of second
order in $K_{s}$ arises in the particle effective energy.
\cite{garkac03prl} The spin noncollinearity depends on the
orientation of $\mathbf{m}$ and results in the effective cubic
anisotropy
\begin{equation}
\mathcal{E}_{2}=K_{\mathrm{ca}}\left(m_{x}^{4}+m_{y}^{4}+m_{z}^{4}\right) ,\qquad K_{\mathrm{ca}}\sim \kappa \frac{K_{s}^{2}}{zJ},  \label{eq:4thorderEn}
\end{equation}
where $z$ is the number of nearest neighbors and for the sc
lattice one has $\kappa \simeq 0.53466.$ This equation was
obtained analytically for $K_{s}\ll J$ in the range of particle
sizes large enough ($\mathcal{N} \gg 1$) but small enough so that
$\delta \psi $ remains small. Numerical calculations yield
$K_{\mathrm{ca}}$ slightly dependent on the size since the
applicability conditions for Eq. (\ref{eq:4thorderEn}) are usually
not fully satisfied. The ratio of the second- to first-order
surface contributions is
\begin{equation}
\frac{\mathcal{E}_{2}}{\mathcal{E}_{1}}\sim \frac{K_{s}}{J}\frac{\mathcal{N}^{1/3}}{\epsilon }.
\label{E2E1Ratio}
\end{equation}
It can be significant even for $K_{s}\ll J$ due to the combined
influence of the large particle size and small deviation from
symmetry, $\epsilon \ll 1. $ Since $K_{\mathrm{ca}}$ is nearly
size independent (i.e., the whole energy of the particle scales
with the volume), it is difficult to experimentally distinguish
between the core cubic anisotropy and that due to the second-order
surface contribution (see discussion later on). The reason for the
size independence of $K_{\mathrm{ca}}$ is the deep penetration of
spin noncollinearities into the core of the particle. This means
that the angular dependence of the noncollinear state also
contributes to the effective anisotropy. Interestingly this
implies that the influence of the surface anisotropy on the
overall effective anisotropy is not an isolated surface phenomena
and is dependent on the magnetic state of the particle. We note
that this effect is quenched by the presence of the core
anisotropy which could screen the effect at a distance of the
order of domain wall width from the surface.

Taking into account the core anisotropy analytically to describe
corrections to Eq. (\ref{eq:4thorderEn}) due to the screening of
spin noncollinearities in the general case is difficult. However,
one can consider this effect perturbatively, at least to clarify
the validity limits of Eq. (\ref{eq:4thorderEn}). One obtains
\cite{kacgar06prep} an additional mixed contribution that is
second order in $K_{s}$ and first order in $K_{c}$
\begin{equation}
\mathcal{E}_{21}=K_{\mathrm{csm}}\,g(\mathbf{m}),\qquad
K_{\mathrm{csm}}\sim
\tilde{\kappa}N_{s}\frac{K_{c}K_{s}^{2}}{J^{2}}  \label{eq:CSM}
\end{equation}
where $g(\mathbf{m})$ is a function of $m_{\alpha }$ which
comprises, among other contributions, both the $2^{\mathrm{nd}}$-
and $4^{\mathrm{th}}$-order contributions in spin components
\cite{kacgar06prep}. For example for sc lattice,
$g(\theta,\varphi=0) = -\cos^2\theta + 3\cos^4\theta -
2\cos^6\theta$, which is shown later to give agreement with
numerical simulations. This mixed contribution, called here the
\textit{core-surface mixing} (CSM) contribution, should satisfy
$K_{\mathrm{csm}}\lesssim K_{\mathrm{ca}}$ which requires
\begin{equation}
N_{s}K_{c}/J\lesssim 1.  \label{KcCondition}
\end{equation}
This is exactly the condition that the screening length (i.e., the domain-wall width) is still much greater than the linear size of the particle, $\delta \sim\sqrt{J/K_{c}}\gtrsim D\sim \mathcal{N}^{1/3}$. For too large sizes the perturbative treatment becomes invalid.

Thus we have seen that in most cases the effective anisotropy of a magnetic particle, considered as a single magnetic moment, can be approximately described as a combination of uniaxial and cubic anisotropies \cite{garkac03prl, kacbon06prb, kacgar06prep}. Consequently, collecting all these contributions and defining the EOSP energy as
\begin{equation}
\mathcal{E}_{\mathrm{EOSP}}=\frac{1}{J}(\mathcal{E}_c + \mathcal{E}_1 + \mathcal{E}_2 + \mathcal{E}_{21})
\end{equation}
one can model the energy of a many-spin particle as
\begin{equation}\label{eq:FullEOSPEn}
\mathcal{E}_{\mathrm{EOSP}}=-k_{\mathrm{ua}}^{\mathrm{eff}}m_{z}^{2} - \frac{1}{2}k_{\mathrm{ca}}^{\mathrm{eff}}\sum_{\alpha=x,y,z}m_{\alpha }^{4}.
\end{equation}
The subscripts $\mathrm{ua/ca}$ stand for uniaxial/cubic
anisotropy, respectively. The effective anisotropy constants are
normalized to the exchange constant $J$, according to the
definition Eq. (\ref{def}). Note that we have changed the sign of
the cubic anisotropy constant to be consistent with more customary
notations.

We would like to remark here that the effective energy potential
(\ref{eq:FullEOSPEn}) is an approximation to the full energy of a
many-spin particle, especially with respect to CSM contribution
(\ref{eq:CSM}). More precisely, the function $g(\mathbf{m})$ in
(\ref{eq:CSM}) contains terms of various orders (see the
expression in the text et seq.), and thus fitting the many-spin
particle's energy to the effective potential (\ref{eq:FullEOSPEn})
amounts to truncating the function $g(\mathbf{m})$ to a
$4^\mathrm{th}-$order polynomial. As such, when the orders of the
anisotropy contributions of the initial many-particle match those
of the effective potential (\ref{eq:FullEOSPEn}), they get
renormalized by the corresponding contributions from
(\ref{eq:CSM}). Indeed, in Ref.~\onlinecite{kacbon06prb}, where
the energy of a many-spin spherical particle with uniaxial
anisotropy in the core and TSA on the surface was computed using
the Lagrange-multiplier technique, it turned out that the core
anisotropy is modified. However,  when there is no uniaxial
anisotropy in the initial many-spin particle, i.e., when the core
anisotropy is cubic and the particle is perfectly symmetrical (no
elongation), truncating the function $g(\mathbf{m})$ to the
$4^\mathrm{th}-$order generates an artificial uniaxial anisotropy,
very small though. On the other hand, even if the core anisotropy
is not uniaxial but the particle presents some elongation, the
effective energy does exhibit a relatively large uniaxial
contribution induced by the surface due to the term in Eq.
(\ref{eq:EnElong}). Therefore, the $2^{\mathrm{nd}}$-order term
with the coefficient $k_{\mathrm{ua}}^{\mathrm{eff}}$ in Eq.
(\ref{eq:FullEOSPEn}) stems, in general, from the two
contributions (\ref{eq:EnElong}) and (\ref{eq:CSM}). Similarly,
the $4^{\mathrm{th}}$-order term with the coefficient
$k_{\mathrm{ca}}^{\mathrm{eff}}$ comprises the contribution
(\ref{eq:4thorderEn}) from the surface,  and part of the CSM
contribution in Eq. (\ref{eq:CSM}), and also a contribution from
the core if the latter has a cubic anisotropy.

We now are going to numerically calculate the effective energy of spherical, ellipsoidal and truncated octahedral magnetic particles cut from a lattice with sc, fcc, and hcp structures. We will plot this energy as a function of the polar angles of the net magnetic moment $\mathbf{m}$ of the particle and fit it to Eq. (\ref{eq:FullEOSPEn}). From these fits we extract the effective anisotropy constants $k_{\mathrm{ua}}^{\mathrm{eff}}$ and $k_{\mathrm{ca}}^{\mathrm{eff}}$ and compare their behavior with those predicted by the analytical formulas discussed above in the case of the sc lattice. We will investigate the differences between the results for different lattice
structures and crystal shapes.

\section{Numerical method and results}

\label{Sec-Numerical}

\subsection{\label{sec:ComMethod}Computing method}

As mentioned above, the problem of studying the multidimensional energy landscape of the multispin particle
is, in general, very difficult. However, if the exchange $%
J $ is dominant over the anisotropy and the spin noncollinearity
is small, one can, to a good approximation, describe the particle
by its net or global magnetization $\mathbf{m}$ ($\left|
\mathbf{m}\right| =1$) as a slow master variable. All other
variables such as spin noncollinearities quickly adjust themselves
to the instantaneous direction of $\mathbf{m.}$ Thus one can
treat the energy of the particle as a function of $%
\mathbf{m}$ only. Technically this can be done with the help of the Lagrange
multiplier technique \cite{garkac03prl} by considering the augmented energy $%
\mathcal{F}=\mathcal{H}-\mathcal{N}\mathbf{\lambda }\cdot (\mathbf{\nu }-%
\mathbf{m}),$ where $\mathbf{\nu }\equiv \sum_{i}\mathbf{s}_{i}/|\sum_{i}%
\mathbf{s}_{i}|$ and $\mathcal{H}$ is the atomistic energy of the particle, Eq. (%
\ref{eq:Hamiltonian})$.$ The Lagrange-multiplier term produces an additional
torque on the atomic spins that forces the microscopic net magnetization $%
\mathbf{\nu }$ to coincide with $\mathbf{m}$. The equilibrium
state of the spin system is determined by solving the
Landau-Lifshitz equation (without the precession term and with the
damping coefficient $\alpha =1$) and an additional equation for
$\mathbf{\lambda }$
\begin{equation}
\frac{d\mathbf{s}_{i}}{dt}=-\left[ \,\mathbf{s}_{i}\times \left[ \mathbf{s}%
_{i}\times \mathbf{F}_{i}\right] \right] ,\qquad \frac{d\mathbf{\lambda }}{dt%
}=-\mathcal{N}(\mathbf{\nu }-\mathbf{m}),
\end{equation}
where the effective field $\mathbf{F}_{i}=-\partial
\mathcal{F}/\partial \mathbf{s}_{i}$ depends on $\mathbf{\lambda
}$. It is worth noting that the stationary points of $\mathcal{F}$
found with this method are also stationary points of the actual
Hamiltonian $\mathcal{H.}$ Indeed, for these orientations of
$\mathbf{m}$ no additional torque is needed to support this state,
thus the solution of our equations yields $[\mathbf{\lambda
}\times \mathbf{m}]=0$. For all other directions, the unphysical
Lagrange-multiplier field introduces distortions of the
microscopic spin configuration. Nevertheless, if the exchange is
dominant, these distortions remain small. A further advantage of
this technique is that it can produce highly non-collinear
multi-dimensional stationary points \cite
{garkac03prl,kacbon06prb,pazetal06unpub} in the case of strong
surface anisotropy. Here, again, these stationary points are true
points because of the condition $[\mathbf{\lambda }\times
\mathbf{m}]=0.$ In order to check the correct loci of the saddle
points, we computed the eigenvalues and gradients of the Hessian
matrix associated with the Hamiltonian of Eq. (\ref
{eq:Hamiltonian}), with the results that are exactly the same as
those obtained by the much simpler Lagrange-multiplier method. We
stress that previous researchers (see, e.g., Ref.
\onlinecite{jametetal04prb}), ignore the effect of spin
non-collinearities.

In this work we show that the magnetic behavior of small particles is very
sensitive to the surface arrangement, shape of the particles and underlying
crystallographic structure. To investigate the various tendencies, we have
considered particles cut from lattices with the simple cubic (sc),
body-centered cubic (bcc), face-centered cubic (fcc), hexagon closed-packed
(hcp). Although experimental studies providing transmission electron
microscopy images show particles resembling truncated octahedra \cite
{jametetal01prl,jametetal04prb}, making realistic particle shapes and
surface arrangements proves to be rather complex. Truncated octahedra have
been included in our studies as an ideal case for fcc crystals. The reality
is somewhat more complicated, though. In Ref. \onlinecite{jametetal04prb},
in order to interpret the experimental results of the $3D$-dimensional
switching field curve, the so-called Stoner-Wohlfarth astroid, it was
assumed that a few outer layers in the truncated octahedral particle were
magnetically ``dead'', leading to an effective elongation and thereby to a
non-perfect octahedron. Producing such a faceted elongated particle by
somehow cutting the latter is an arbitrary procedure. In order to minimize
the changes in the surface structure caused by elongation, we assumed a
spherical particle or introduced elliptical elongation along the easy axis.
This kind of structure has been the basis of many theoretical studies using
the Heisenberg Hamiltonian (see, e.g., Refs.
\onlinecite{kodber99prb,
kacetal00epjb, igllab01prb, kacdim02prb, kacmah04jmmm, garkac03prl, kacgar05springer, kacbon06prb}).

Regarding the arrangement at the particle surface, an appropriate
approach would be to use molecular-dynamic techniques \cite
{dawbas83prl,dorfbaueretal06jap} based on the empirical potentials
for specific materials. This would produce more realistic
non-perfect surface structures, more representative of what it is
hinted at by experiments. However, these potentials exist only for
some specific materials and do not fully include the complex
character of the surface.
%
\begin{figure}[tbp]
\includegraphics[width=8cm]{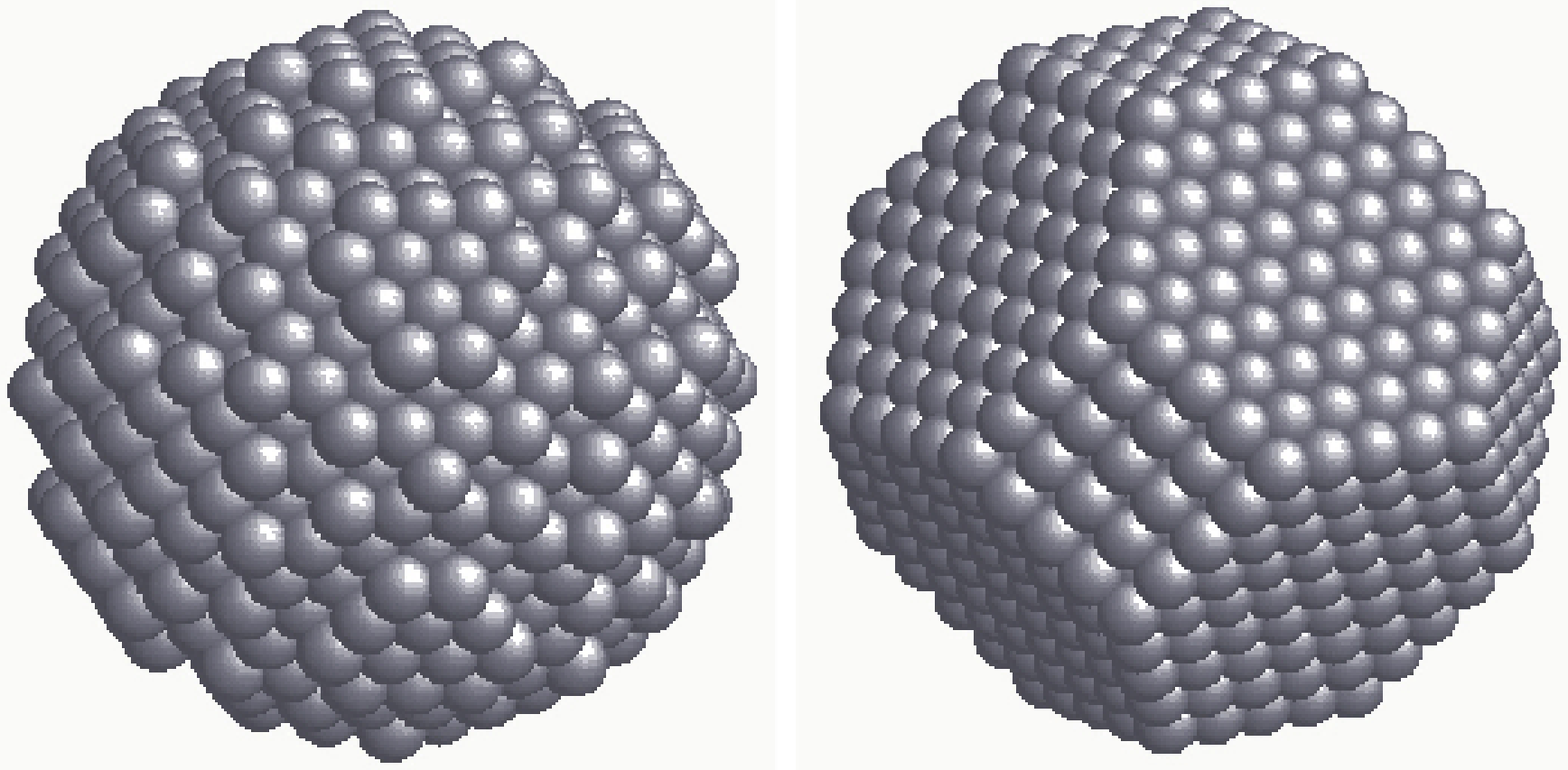}
\caption{Two particles cut from fcc structure: spherical (left)
and truncated octahedron (right).} \label{fig:Twoparticles}
\end{figure}
%
Moreover, the particles thus obtained (see, e.g.,
Ref.~\onlinecite{evansetal06jap}), may have non-symmetric
structures, and may present some dislocations. All these phenomena
lead to a different behavior of differently prepared particles
which will be studied in a separate publication.

In the present work, in order to illustrate the general tendency of the magnetic behavior, we mostly present results for particles with ``pure'' non modified surfaces, namely spheres, ellipsoids and truncated octahedra cut from regular lattices. Even in this case, the surface arrangement may appear to be very different (see Fig. \ref{fig:Twoparticles}) leading to a rich magnetic behavior.

\subsection{\label{subsec:spheres}Spherical particles}

\begin{figure*}[tbp]
\includegraphics[width=7.5cm]{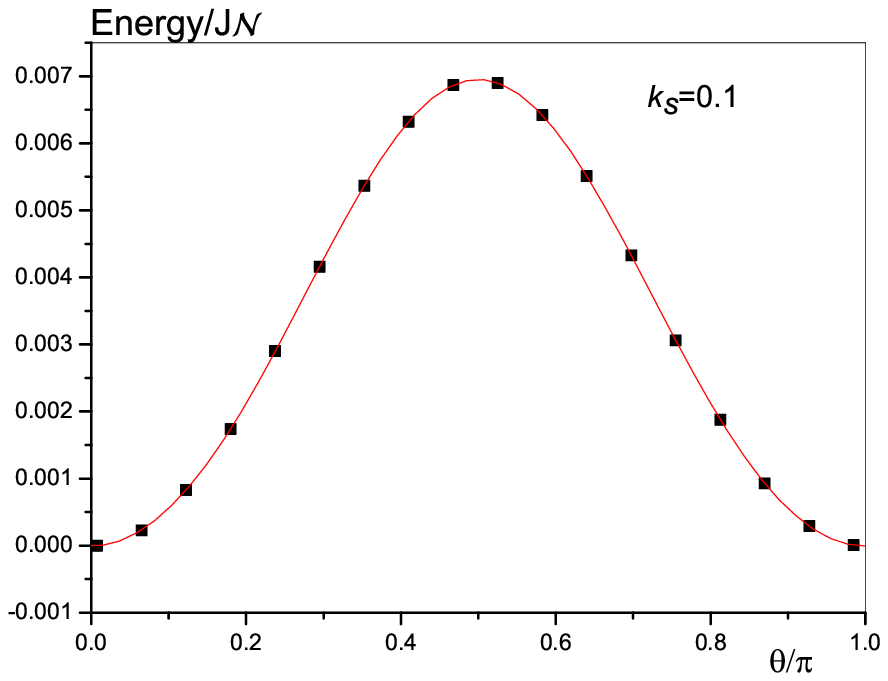} %
\includegraphics[width=7.5cm]{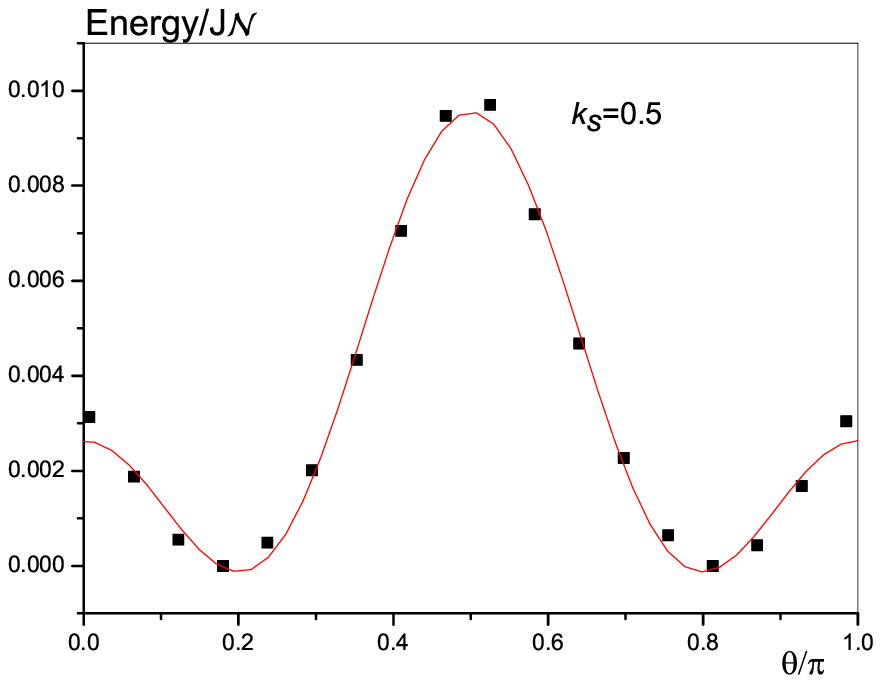}
\caption{(Shifted) $2D$ energy potentials of a spherical many-spin particle
of $\mathcal{N}=1736$ spins on an sc lattice with uniaxial anisotropy in the
core ($k_{c}=0.01$) and NSA with constant $k_{s}=0.1$ (left), $0.5$ (right).
The solid lines are plots of Eq.~(\protect\ref{eq:FullEOSPEn}). }
\label{fig:En2dMSP_sphere_sc_fit_ks01035}
\end{figure*}

We compute the $3D$ energy potential as a function of the spherical coordinates $(\theta, \varphi)$ of the net magnetization $\mathbf{m}$ of a  many-spin particle. We do this for a spherical particle with uniaxial anisotropy in the core and NSA, cut from sc, fcc and hcp lattices, and for different values of the surface anisotropy constant, $k_{s}
$. %
\begin{figure*}[tbp]
\includegraphics[width=9.0cm]{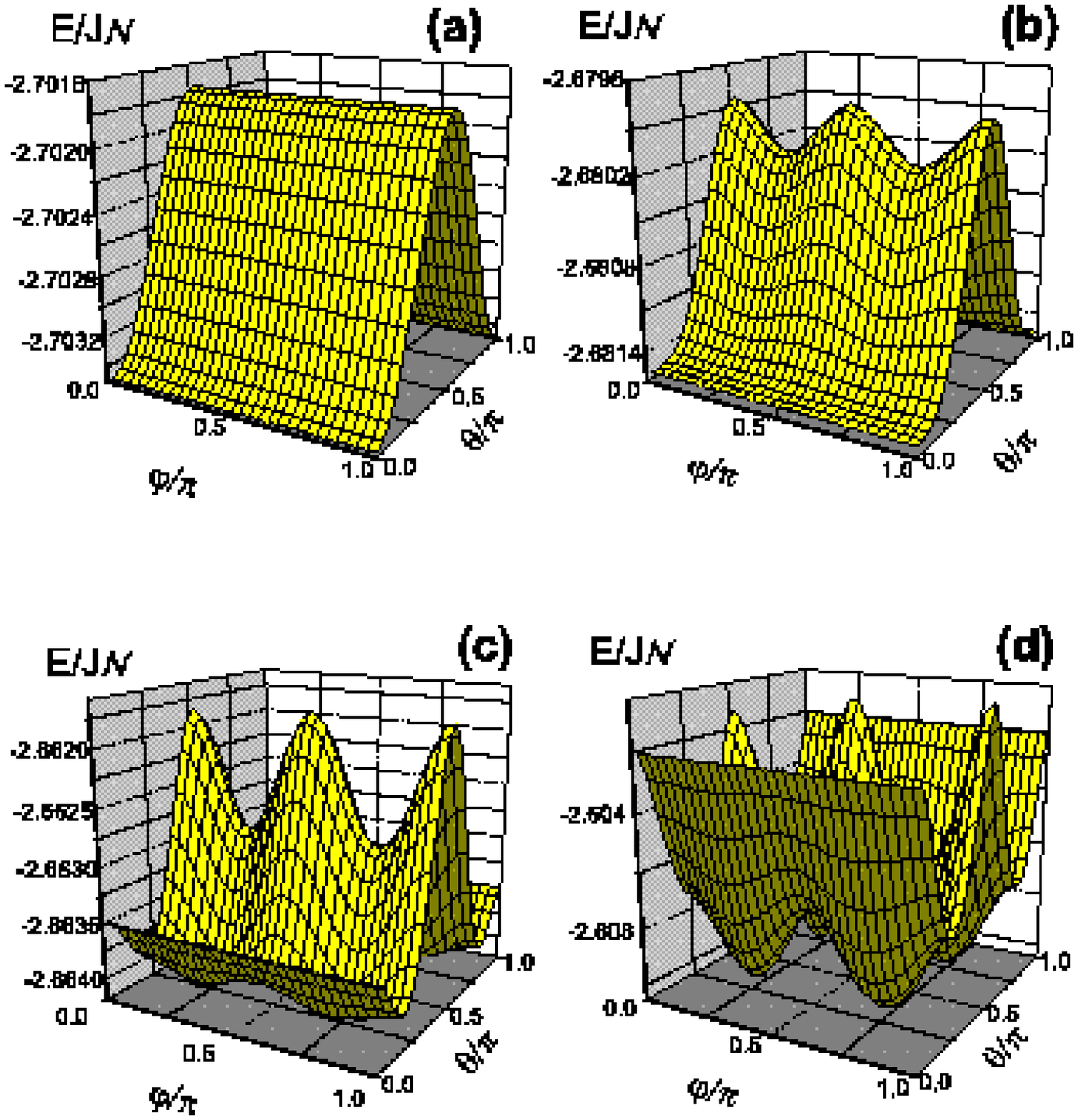}
\caption{Energy potentials of a spherical many-spin particle of $\mathcal{N}=1736$ spins on an sc lattice with uniaxial anisotropy in the core ($k_{c}=0.0025$) and NSA with constant (a) $k_{s}=0.005$, (b) $k_{s}=0.112$, (c) $k_{s}=0.2$, (d) $k_{s}=0.5$. }
\label{fig:EnMSP_sphere_sc}
\end{figure*}
%
Fig.~\ref{fig:EnMSP_sphere_sc} shows energy landscapes for
spherical particles cut from an sc lattice. One can see that as
$k_{s}$ increases the global minima move away from those defined
by the core uniaxial anisotropy, i.e., at $\theta =0,\pi $ and any
$\varphi $, and become maxima, while new minima and saddle points
develop which are reminiscent of cubic anisotropy. Now, in
Fig.~\ref{fig:En2dMSP_sphere_sc_fit_ks01035} we present the
corresponding $2D$ energy potential ($\varphi =0$). From this
graph we see that the energy of the many-spin particle is well
reproduceed by Eq.~(\ref{eq:FullEOSPEn}) when $k_{s}$ is small,
which shows that such a many-spin particle can be treated as an EOSP
with an energy that contains uniaxial and cubic anisotropies.
However, as was shown in Ref.~\onlinecite{kacbon06prb}, when the
surface anisotropy increases to larger values this mapping of
the many-spin particle onto an effective one-spin particle fails.
%
%

\begin{figure*}[tbp]
\includegraphics[width=7.5cm]{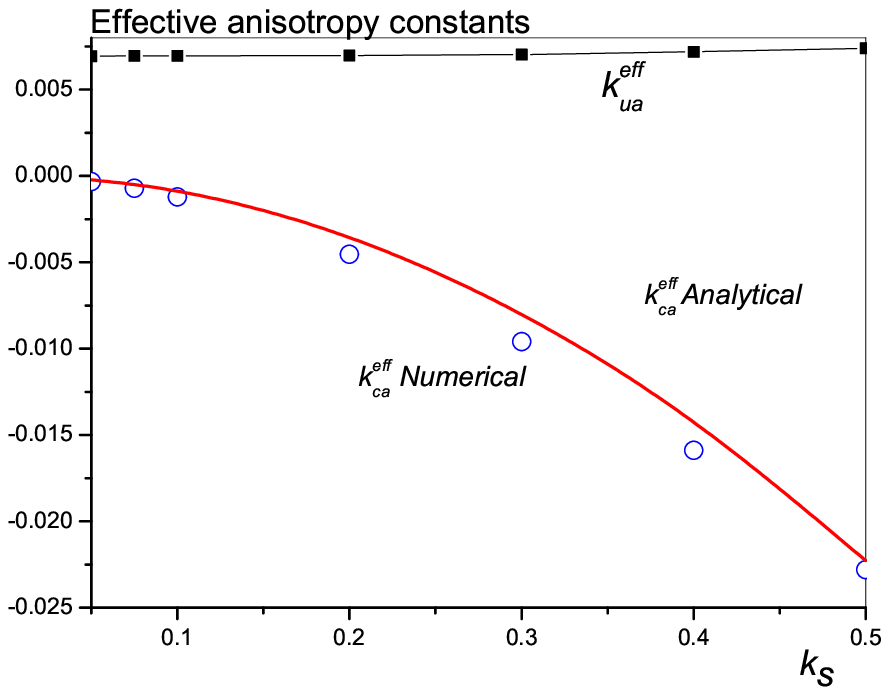}
\includegraphics[width=7.5cm]{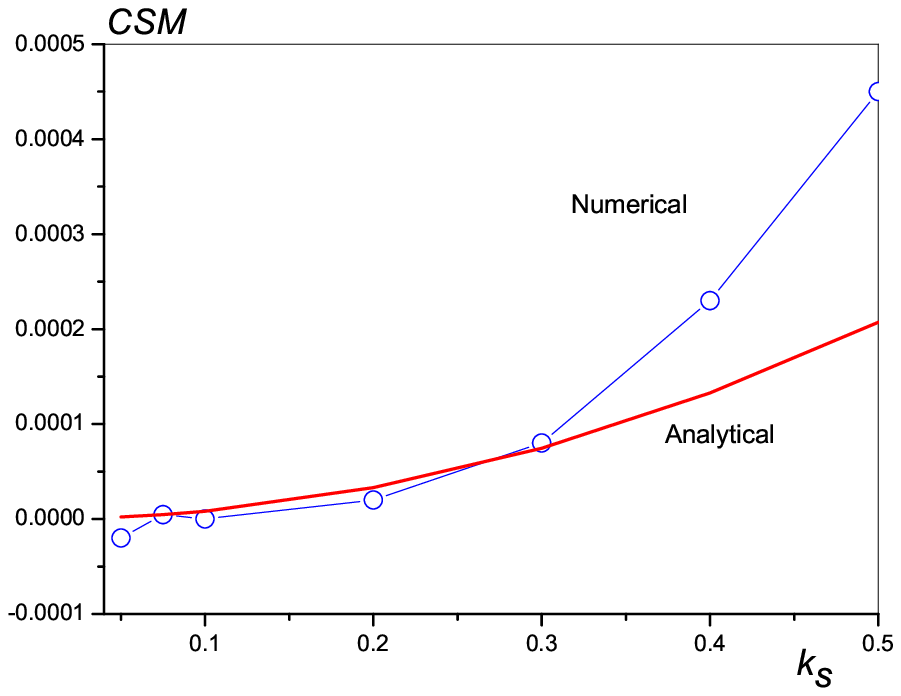}
\caption{Effective anisotropy constants against $k_{s}$ for a
spherical many-spin particle of $\mathcal{N}=1736$ spins cut from
an sc lattice. The panel on the right shows the CSM contribution
obtained numerically as $k_{\mathrm{ua}}^{\mathrm{eff}}$, excluding the core uniaxial contribution from Eq.~(\protect\ref{EcDef}). The thick solid lines are plots of Eqs. (\protect\ref{eq:4thorderEn}) and (\protect\ref{eq:CSM}) with $\varphi=0$.}
\label{fig:KMSP_sphere_sc_fit}
\end{figure*}
%
Repeating this fitting procedure for other values of $k_{s}$ we
obtain the plots of $k_{\mathrm{ua}}^{\mathrm{eff}}$ and
$k_{\mathrm{ca}}^{\mathrm{eff}} $ in
Fig.~\ref{fig:KMSP_sphere_sc_fit}. We first see that these
effective constants are quadratic in $k_{s}$, in accordance with
Eqs.~(\ref{eq:4thorderEn}) and (\ref{eq:CSM}). In addition, the
plot on the right shows an agreement between the constant
$k_{\mathrm{ua}}^{\mathrm{eff}}$, excluding the core uniaxial contribution in Eq.~(\ref{EcDef}). These results confirm those of
Refs.~\onlinecite{kacbon06prb, kacgar06prep, kachkachi06SS} that
the core anisotropy is modified by the
surface anisotropy, though only slightly in the present case.
%
\begin{figure*}[tbp]
\includegraphics[width=9.0cm]{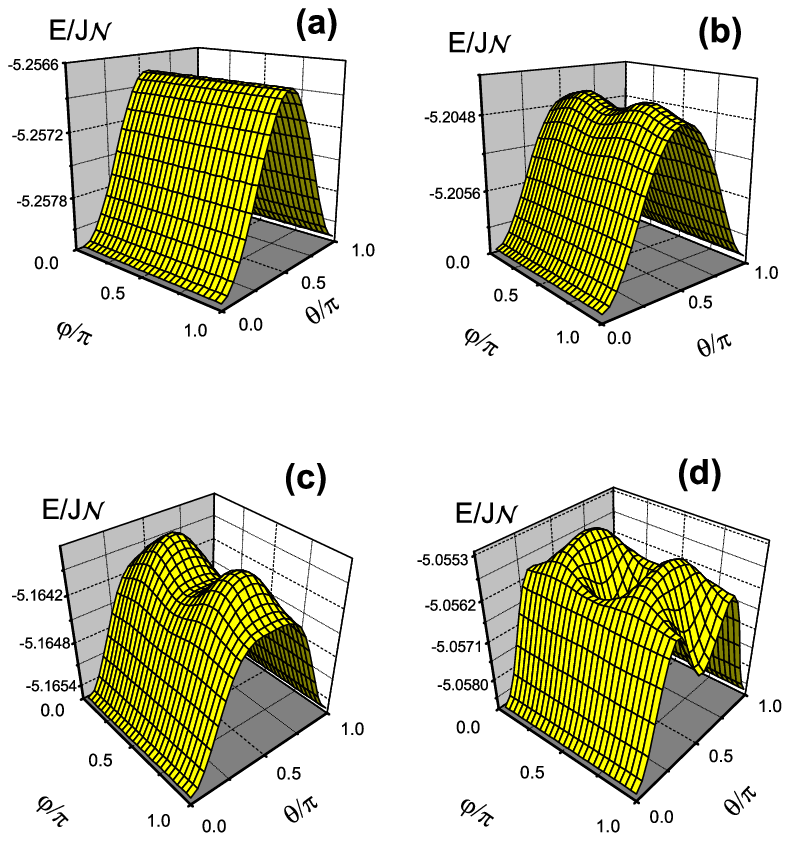}
\caption{Energy potentials of a spherical many-spin particle with
uniaxial anisotropy in the core ($k_{c}=0.0025$) and NSA with
constant (a) $k_{s}=0.005$, (b)$k_{s}=0.1$, (c) $k_{s}=0.175$, (d)
$k_{s}=0.375$. The particle contains $\mathcal{N}=1264$ spins on
an fcc lattice.} \label{fig:EnMSP_sphere_fcc}
\end{figure*}
%

\begin{figure*}[tbp]
\includegraphics[width=7.5cm]{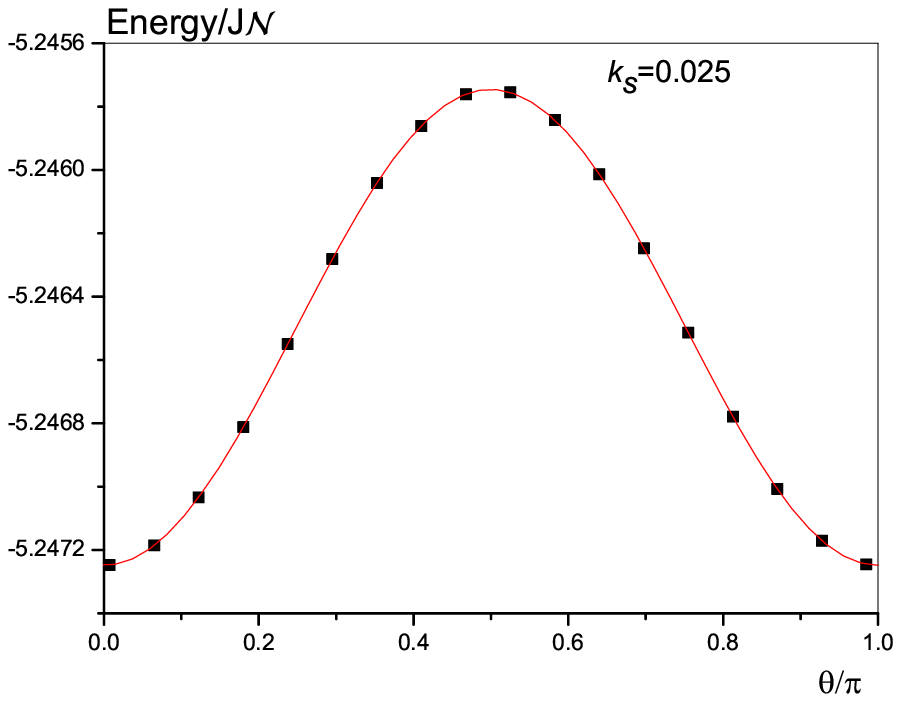}
\includegraphics[width=7.5cm]{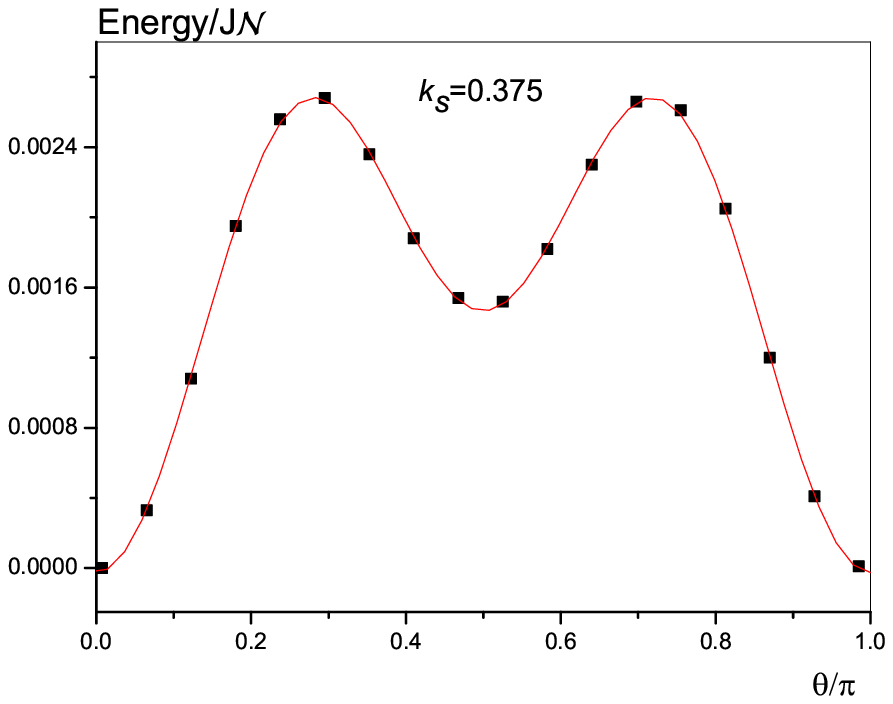}
\caption{The same as in Fig.~\protect\ref{fig:En2dMSP_sphere_sc_fit_ks01035} but here for the fcc lattice and $k_{s}=0.025,0.375$. }
\label{fig:En2dMSP_sphere_fcc_fit_ks01035}
\end{figure*}
%
Comparing the energy potential in Fig.~\ref{fig:EnMSP_sphere_sc}
for the sc lattice and Fig.~\ref{fig:EnMSP_sphere_fcc} for the fcc lattice
one realizes that, because of the fact that different underlying
structure produces different surface spin arrangements, the
corresponding energy potentials exhibit different topologies. For
instance, it can be seen that the point $\theta =\pi /2,\varphi
=\pi /4$ is a saddle in particles cut from an sc lattice and a
maximum in those cut from the fcc lattice. %
\begin{figure}[tbp]
\includegraphics[width=7.5cm]{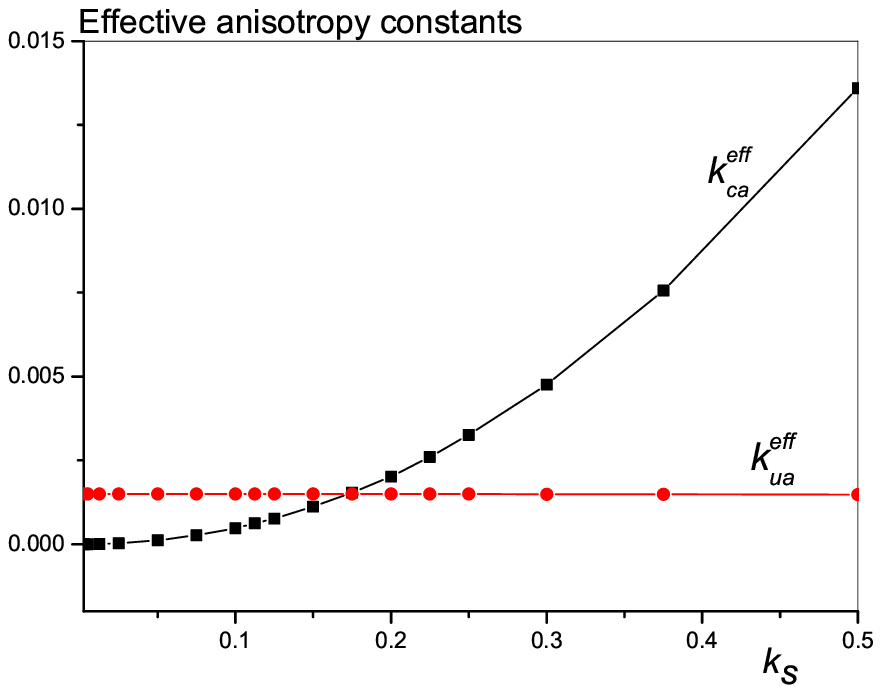}
\caption{Effective anisotropy constants against $k_{s}$ for a spherical particle of $\mathcal{N}=1264$ spins cut from an fcc lattice with uniaxial core anisotropy $k_{c}=0.0025$. The lines are guides to the eye.}
\label{fig:KMSP_sphere_fcc_fit}
\end{figure}
%
Spherical particles cut from the sc lattice exhibit an effective
four-fold anisotropy with $k_{\mathrm{ca}}^{\mathrm{eff}}<0$ [see
Fig.~\ref{fig:KMSP_sphere_sc_fit} and Eq.~(\ref{eq:FullEOSPEn})].
As such, the contribution of the latter to the effective energy is
positive, and this is compatible with the sign of
$K_{\mathrm{ca}}$ in Eq.~(\ref {eq:4thorderEn}). In
Figs.~\ref{fig:En2dMSP_sphere_fcc_fit_ks01035},
\ref{fig:KMSP_sphere_fcc_fit} we plot the $2D$ energy potential
and the effective anisotropy constants, respectively, for a
spherical particle with fcc structure. For a spherical fcc
particle, the effective cubic constant
$k_{\mathrm{ca}}^{\mathrm{eff}}$ is positive [see
Fig.~\ref{fig:KMSP_sphere_fcc_fit}], and as for the sc lattice, it
is quadratic in $k_{s}$. As mentioned earlier, the coefficient
$\kappa $ in (\ref{eq:4thorderEn}) depends on the lattice
structure and for fcc it may become negative. To check this one
first has to find an analytical expression for the spin density on
the fcc lattice, in the same way the sc lattice density was
obtained in Ref.~\onlinecite{garkac03prl} [see Eq.~(6) therein].
The corresponding developments are somewhat cumbersome and are now
in progress. Likewise, the coefficient $\tilde{\kappa}$ in
Eq.~(\ref{eq:CSM}) should change on the fcc lattice, thus changing
the uniaxial and cubic contributions as well.

Finally, in particles with the hcp lattice and large surface
anisotropy, we have found that the effective energy potential is
six-fold, owing to the six-fold symmetry inherent to the hcp
crystal structure. The global magnetization minimum is also
shifted away from the core easy direction.

\subsection{\label{subsec:ellipsoids}Ellipsoidal many-spin particles and effect of elongation}

Now we investigate the effect of elongation. As discussed earlier, due to
the contribution in Eq.~(\ref{eq:EnElong}), even a small elongation may have
a strong effect on the energy barrier of the many-spin particle, and in
particular on the effective uniaxial constant $k_\mathrm{ua}^\mathrm{eff}$,
as will be seen below. Fig.~\ref{fig:EnMSP_elipse_fcc} shows the energy
potential of an ellipsoidal many-spin particle with aspect ratio 2:3, cut
from an fcc lattice. Unlike the energy potentials of spherical particles, the
result here shows that for large surface anisotropy the energy minimum
corresponds to $\theta=\pi/2$. Indeed, due to a large number of local easy
axes on the surface pointing perpendicular to the core easy axis, the total
effect is to change this point from a saddle for small $k_s$ to a minimum
when $k_s$ assumes large values. %
\begin{figure*}[]
\includegraphics[width=9.0cm]{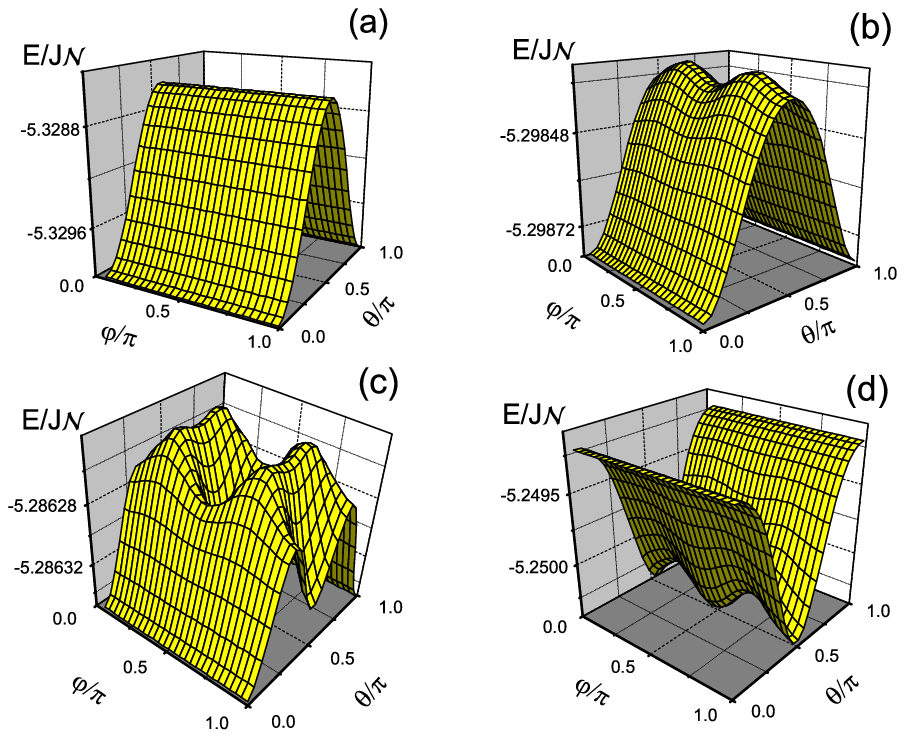}
\caption{Energy potentials of an ellipsoidal particle cut from an fcc
lattice and with uniaxial anisotropy in the core ($k_c = 0.0025$) and NSA
with constant (a) $k_s = 0.0125$, (b)$k_s = 0.075$, (c) $k_s = 0.1$, (d) $k_s = 0.175$.}
\label{fig:EnMSP_elipse_fcc}
\end{figure*}
%
\begin{figure}[]
\includegraphics[width=8cm]{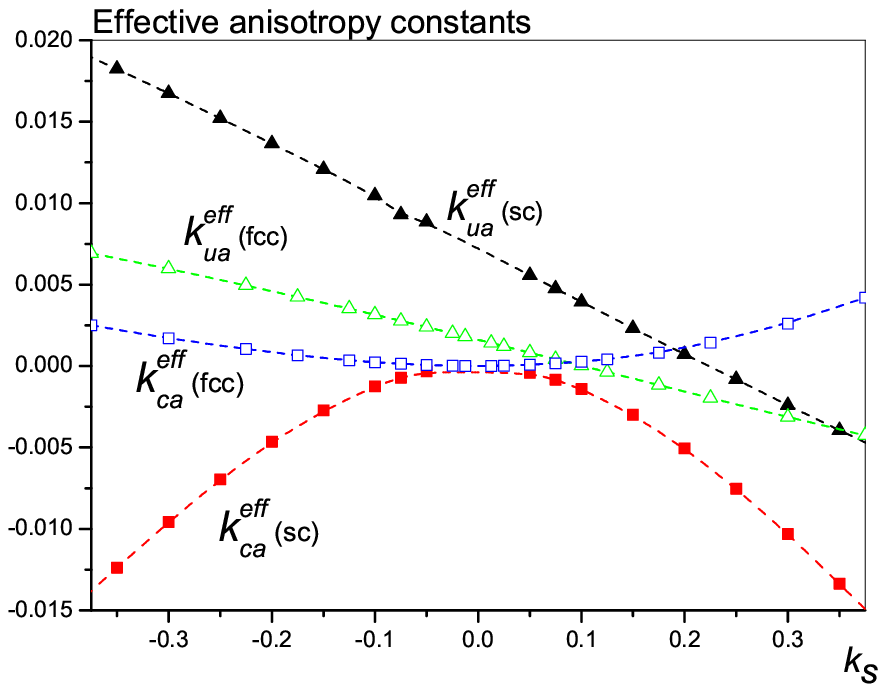}
\caption{Effective anisotropy constants against $k_s$ for an ellipsoidal
particle of $\mathcal{N} = 2044$ spins on sc and fcc lattices, with uniaxial
core anisotropy $k_c=0.0025$. The lines are guides to the eye.}
\label{fig:KMSP_ellipse_fcc_ua_fit}
\end{figure}
%
The effective uniaxial and cubic anisotropy constants are shown in
Fig.~\ref {fig:KMSP_ellipse_fcc_ua_fit}. As expected the effective
uniaxial constant shows a strong linear variation and even changes
sign at some value of $k_s$, as opposed to the case of a spherical
many-spin particle. On the other hand, as for the latter case, the
constant $k_\mathrm{ca}^\mathrm{eff}$ retains
its behavior as a quadratic function of $k_s$. Again, in the case of an sc lattice $k_\mathrm{ca}^\mathrm{eff}<0$ and on an fcc lattice $k_\mathrm{ca}^\mathrm{eff}>0$.

\subsection{\label{subsec:cubooctah}Truncated octahedral many-spin particles}

Here we consider the so-called truncated octahedral particles as
an example of a minimum close-packed cluster structure.  Real
particles are often reported as having this structure with fcc or
bcc underlying lattice [see, e.g. Co or Co/Ag particles in
Ref.~\onlinecite{jametetal01prl, jametetal04prb, luisetal02prb}].

\begin{figure}[]
\includegraphics[width=8cm]{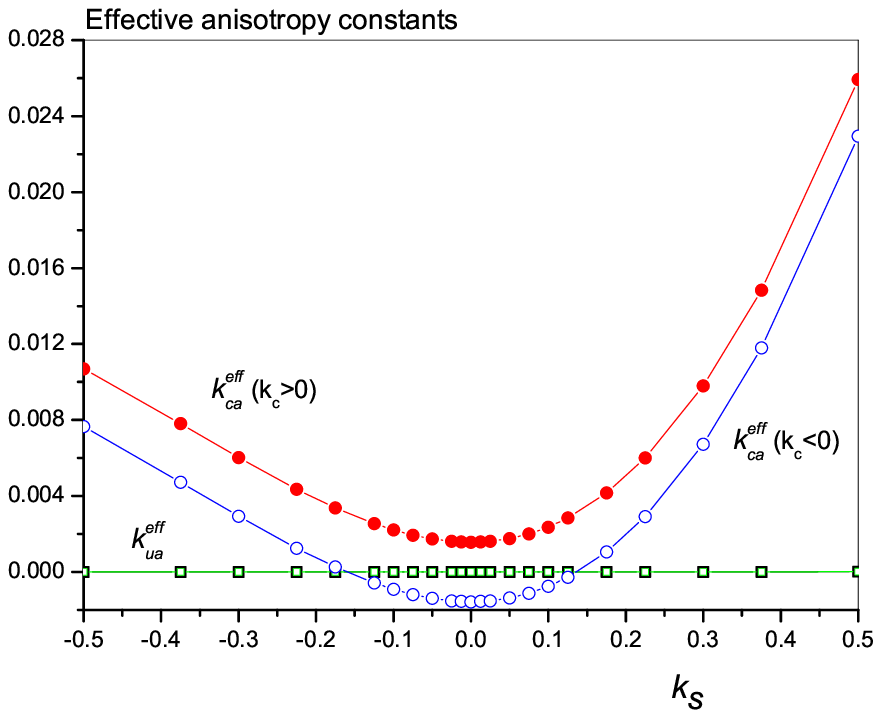}
\caption{Effective anisotropy constants against $k_s$ for a
regular truncated octahedral particle of $\mathcal{N} = 1289$
spins, fcc structure and cubic anisotropy in the core with $k_c>0$
and $k_c<0$.} \label{fig:KMSP_cubooctah_fcc_ca_fit1}
\end{figure}

Regular truncated octahedral having six squares and eight hexagons
on the surface has been constructed cutting an ideal fcc lattice
in an octahedral (two equal mutually perpendicular pyramids with
square bases parallel to $XY$ plane) and subsequent truncation.
Equal surface densities in all hexagons and squares could be
obtained if the fcc lattice is initially rotated 45 degrees in the
$XY$ plane (the $X$-axis will be parallel to (1,1,0) direction and
the $Z$-axis to (0,0,1) direction). We perform the same
calculations as before for a many-spin particle cut from an fcc
lattice, with cubic single-site anisotropy in the core and NSA.
The results presented in Fig.\ref{fig:KMSP_cubooctah_fcc_ca_fit1}
show zero uniaxial contribution and the modification of the cubic
anisotropy by surface effects. It can be seen that, similarly to
the results discussed above, the effective cubic constant is again
proportional to $k_s^2 $ for small $k_s$. This is mainly due to
the two contributions, one coming from the initial core cubic
anisotropy and the other from the surface contribution as in
Eq.~(\ref{eq:4thorderEn}). It is interesting to note that the
surface contribution can  change the sign of the initially
negative cubic core anisotropy constant. We can also observe an
asymmetric behavior of effective anisotropy constants with respect
to the change of the sign of the surface anisotropy which
 we found, in general, in all particles with fcc underlying
 lattice.

\begin{figure}[]
\includegraphics[width=8cm]{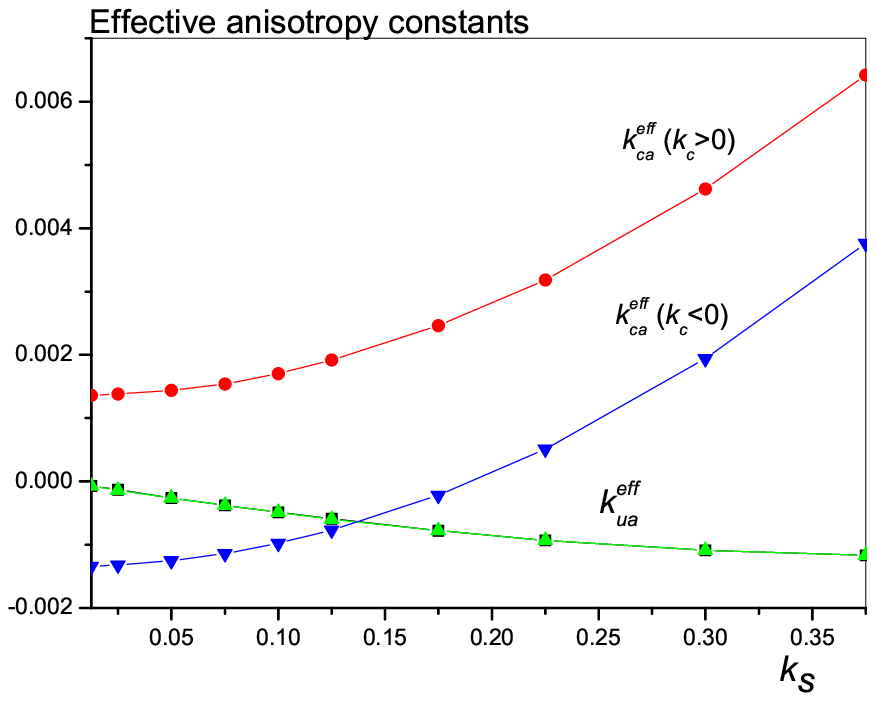}
\caption{Effective anisotropy constants against $k_s$ for a
truncated octahedral particle with different atomic densities on
surfaces. The particle has $\mathcal{N} = 1080$ spins, fcc
structure and cubic anisotropy in the core with $k_c>0$ and
$k_c<0$.} \label{fig:KMSP_cubooctah_fcc_ca_fit}
\end{figure}

%
If the fcc lattice is initially oriented with crystallographic
lattice axes parallel to that of the system of coordinates, then
different atomic densities  are created on different surfaces.
This way the surface density along the $XY$ circumference is
different to that along $XZ$ one. In
Fig.~\ref{fig:KMSP_cubooctah_fcc_ca_fit} we plot the dependence of
the effective cubic constant $k_\mathrm{ca}^\mathrm{eff}$ as a
function of $k_s$ for $k_c>0$ and $k_c<0$.  The surface
contribution can again change the sign of the initially negative
cubic core anisotropy constant. Besides, we clearly see that the
many-spin particle develops a negative uniaxial anisotropy
contribution, induced by the surface in the presence of core
anisotropy. This constitutes a clear example of importance of the
surface arrangement for global magnetic properties of individual
particles.
%
%

\section{Energy barriers}

\subsection{Dependence of the energy barrier on $k_s$}

\begin{figure}[tbp]
\includegraphics[width=7.5cm]{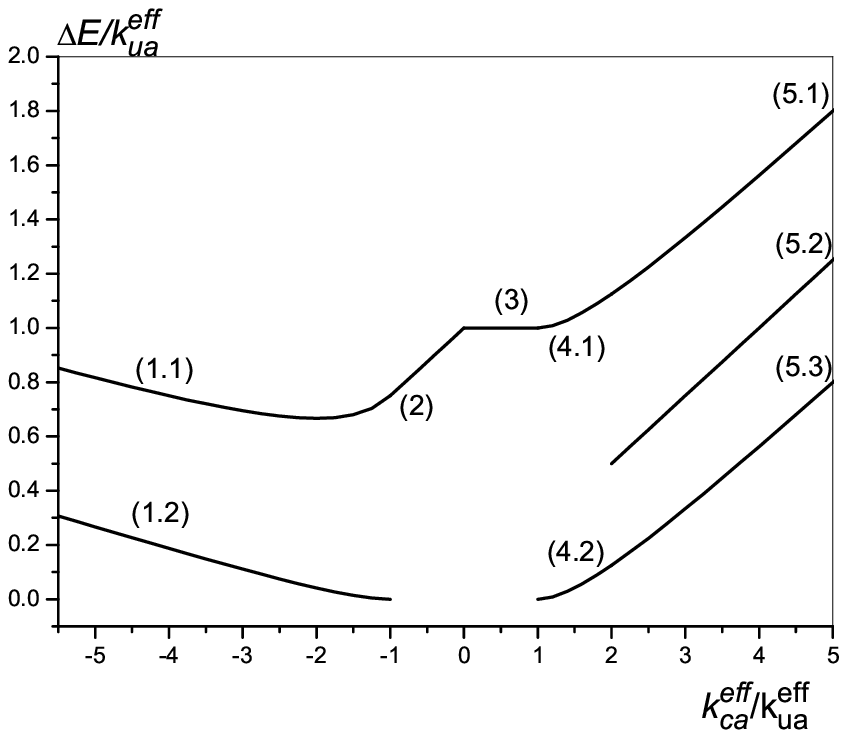}
\caption{Analytical energy barriers of the EOSP with the potential (%
\protect
\ref{eq:FullEOSPEn}) as functions of $k_{\mathrm{ca}}^{\mathrm{eff}}/k_{\mathrm{ua}}^{\mathrm{eff}}=\protect\zeta $ with $k_{\mathrm{ua}}^{\mathrm{eff}}>0$. Analytical formulas in Table I are labelled.}
\label{Bar_OSP}
\end{figure}

Now we evaluate the energy barriers of many-spin particles by
numerically computing the difference between the energy at the
saddle point and at the minimum, using the Lagrangian multiplier
technique described earlier. On the other hand, the EOSP energy
potential (\ref{eq:FullEOSPEn}) can be used to analytically
evaluate such energy barriers and to compare them with their
numerical counterparts. Namely, we have investigated minima,
maxima and saddle points of the effective potential
(\ref{eq:FullEOSPEn}) for different values and signs of the
parameters $k_\mathrm{ua}^\mathrm{eff}$ and
$k_\mathrm{ca}^\mathrm{eff}$ and calculated analytically the
energy barrier in each case. The results are presented in Table I.
The energy barriers for the case $k_\mathrm{ua}^\mathrm{eff} > 0$
are plotted in Fig. \ref{Bar_OSP}. We remark that for large
surface anisotropy $|\zeta| >> 1$, where $ \zeta\equiv
k_\mathrm{ca}^\mathrm{eff}/k_\mathrm{ua}^\mathrm{eff}$, all energy barriers are simple linear
combinations of the two effective anisotropy constants.

\begin{table*}[tbp]
\caption{Energy barriers for the effective one-spin particle. The critical angle
$\protect\theta_c(\protect\varphi)$ is defined by $\cos(\protect\theta_c(\protect\varphi))^2=(k_\mathrm{ua}^\mathrm{eff}+k_\mathrm{ua}^\mathrm{eff}(\sin(\protect\varphi)^4+\cos(\protect\varphi)^4))/(k_\mathrm{ca}^\mathrm{eff}(1+\sin(\protect\varphi)^4+\cos(\protect\varphi)^4))$}
\label{tab:table3}
\begin{ruledtabular}
\begin{tabular}{ccccc}
  & & $k_\mathrm{ua}^\mathrm{eff} >0$ & \\
$\zeta=k_\mathrm{ca}^\mathrm{eff}/k_\mathrm{ua}^\mathrm{eff}$  & $\mathrm{Minima} (\theta,\varphi)$ & Saddle points$(\theta,\varphi)$ & Energy barriers, $\Delta E_{\mathrm{EOSP}} $\\

  \hline

 $-\infty<\zeta<-1$&$\theta_c(\pi/4);\pi/4$ &$\pi/2;\pi/4$&$k_\mathrm{ua}^\mathrm{eff}/3-k_\mathrm{ca}^\mathrm{eff}/12-(k_\mathrm{ua}^\mathrm{eff})^2/3k_\mathrm{ca}^\mathrm{eff}  $ &  (1.1) \\
  & $\theta_c(\pi/4);\pi/4$ &$\theta_c(\pi/2);\pi/2$&$-k_\mathrm{ua}^\mathrm{eff}/6-k_\mathrm{ca}^\mathrm{eff}/12-(
  k_\mathrm{ua}^\mathrm{eff})^2/12k_\mathrm{ca}^\mathrm{eff}$    &      (1.2) \\

\hline
$-1<\zeta<0$&$0;0$ &$\pi/2;0$&$k_\mathrm{ua}^\mathrm{eff}+k_\mathrm{ca}^\mathrm{eff}/4$   &   (2)\\

\hline

$0<\zeta<1$&$0;\pi/2$ &$\pi/2;\pi/2$&$k_\mathrm{ua}^\mathrm{eff}$  &    (3) \\

\hline

$1<\zeta<2$&$\theta_c(0);\pi/2$ &$\theta_c(\pi/2);\pi/2$&$k_\mathrm{ua}^\mathrm{eff}/2+k_\mathrm{ca}^\mathrm{eff}/4+(k_\mathrm{ua}^\mathrm{eff})^2/4k_\mathrm{ca}^\mathrm{eff}  $ &    (4.1) \\
 &$\pi/2;\pi/2$ &$\theta_c(\pi/2);\pi/2$&$-k_\mathrm{ua}^\mathrm{eff}/2+k_\mathrm{ca}^\mathrm{eff}/4+(k_\mathrm{ua}^\mathrm{eff})^2/4k_\mathrm{ca}^\mathrm{eff}     $   &    (4.2) \\

\hline

 $2<\zeta<\infty$&$0;\pi/2$ &$\theta_c(\pi/2);\pi/2$&$k_\mathrm{ua}^\mathrm{eff}/2+k_\mathrm{ca}^\mathrm{eff}/4+(k_\mathrm{ua}^\mathrm{eff})^2/4k_\mathrm{ca}^\mathrm{eff}  $  &    (5.1) \\
  & $\pi/2;\pi/2$ &$\pi/2;\pi/4$&$k_\mathrm{ca}^\mathrm{eff}/4   $ &     (5.2) \\
  & $\pi/2;\pi/2$ &$\theta_c(\pi/2);\pi/2$&$-k_\mathrm{ua}^\mathrm{eff}/2+k_\mathrm{ca}^\mathrm{eff}/4+(k_\mathrm{ua}^\mathrm{eff})^2/4k_\mathrm{ca}^\mathrm{eff}  $    &  (5.3) \\
\hline \hline
\\ & & $k_\mathrm{ua}^\mathrm{eff} <0$ & \\

\hline

 $-\infty<\zeta<-1$&$\pi/2;\pi/2$ &$\theta_c(\pi/2);\pi/2$&$-k_\mathrm{ua}^\mathrm{eff}/2+
 k_\mathrm{ca}^\mathrm{eff}/4+(k_\mathrm{ua}^\mathrm{eff})^2/4k_\mathrm{ca}^\mathrm{eff} $&      (6.1) \\
  & $\pi/2;0$ &$\pi/2;\pi/4$&$k_\mathrm{ca}^\mathrm{eff}/4$&      (6.2) \\
& $0;\pi/2$ &$\theta_c(\pi/2);\pi/4$&$k_\mathrm{ua}^\mathrm{eff}/2+k_\mathrm{ca}^\mathrm{eff}/4+
(k_\mathrm{ua}^\mathrm{eff})^2/4k_\mathrm{ca}^\mathrm{eff}$&      (6.3) \\

\hline
$-1<\zeta<0$&$\pi/2;0$ &$\pi/2;\pi/4$&$k_\mathrm{ca}^\mathrm{eff}/4$ &      (7) \\

\hline

$0<\zeta<1$&$\pi/2;\pi/4$ &$\pi/2;\pi/2$&$k_\mathrm{ca}^\mathrm{eff}/4$ &      (8)  \\

\hline

$1<\zeta<2$&$\pi/2;\pi/4$ &$\theta_c(\pi/2);\pi/2$&$-k_\mathrm{ua}^\mathrm{eff}/2+
(k_\mathrm{ua}^\mathrm{eff})^2/4k_\mathrm{ca}^\mathrm{eff}  $&      (9) \\

\hline

 $2<\zeta<\infty$&$\theta_c(\pi/4);\pi/4$ &$\pi/2;\pi/4$&$ -k_\mathrm{ua}^\mathrm{eff}/6-k_\mathrm{ca}^\mathrm{eff}/12-(k_\mathrm{ua}^\mathrm{eff})^2/3k_\mathrm{ca}^\mathrm{eff}$
 &      (10.1) \\
  & $\theta_c(\pi/4);\pi/4$ &$\theta_c(\pi/2);\pi/2$&$k_\mathrm{ua}^\mathrm{eff}/3 -
  k_\mathrm{ca}^\mathrm{eff}/12 -(k_\mathrm{ua}^\mathrm{eff})^2/12k_\mathrm{ca}^\mathrm{eff} $&      (10.2) \\

\end{tabular}
\end{ruledtabular}
\end{table*}

Note that in a wide range of the parameters several energy
barriers (corresponding to different paths of magnetization
rotation) coexist in the system in accordance with the complex
character of the effective potential with two competing
anisotropies [see Figs.~\ref{fig:EnMSP_sphere_sc}(c),
\ref{fig:EnMSP_elipse_fcc}(c)]. Because of this competition the
symmetry of the anisotropy can be changed leading to relevant
energy barriers in the $\theta$ or $\varphi$ direction, the former
case is illustrated in Fig.~\ref{fig:EnMSP_elipse_fcc}(a,b,c),
where switching occurs between global minima at $\theta=0$ and
$\theta=\pi$ and the latter is shown in
Fig.~\ref{fig:EnMSP_elipse_fcc}(d) where the large surface
anisotropy has given rise to an easy plane with the stable states
corresponding to ($\theta=\pi/2, \varphi=n\times\pi/2$) where $n$ is an
integer. In some cases there are multiple energy barriers, but
here we consider only the relevant energy barrier for switching,
corresponding to the lowest energy path between global minima.

We have seen that in the case of a spherical particle cut from an sc lattice, and in accordance with the EOSP energy potential (\ref {eq:FullEOSPEn}), $k_\mathrm{ua}^\mathrm{eff} > 0$ and $k_\mathrm{ca}^\mathrm{eff} < 0$ [see Fig.~\ref{fig:KMSP_sphere_sc_fit}]. For a spherical particle with an fcc lattice, $k_\mathrm{ua}^\mathrm{eff} > 0, k_\mathrm{ca}^\mathrm{eff} > 0$, as can be seen in Fig.~\ref{fig:KMSP_sphere_fcc_fit}. Finally, for ellipsoidal and truncated octahedral many-spin particles, the results in Figs.~\ref{fig:KMSP_ellipse_fcc_ua_fit}, \ref{fig:KMSP_cubooctah_fcc_ca_fit} show that $k_\mathrm{ua}^\mathrm{eff}$ may become negative at some value of $k_s$, since then the contributions similar to (\ref{eq:EnElong}) and (\ref{eq:CSM}) become important.

%

\begin{figure}[tbp]
\includegraphics[width=7.5cm]{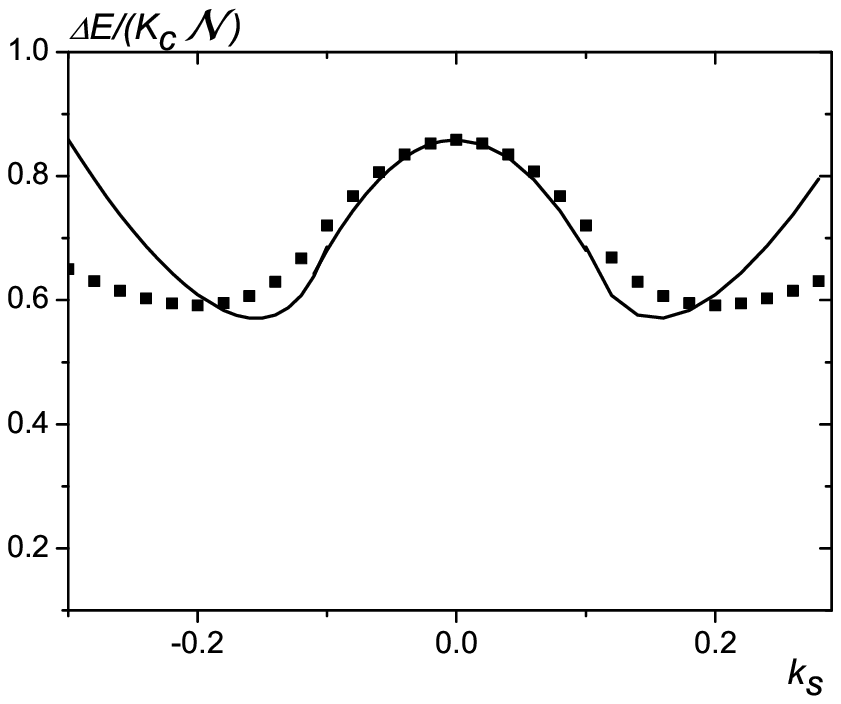}
\caption{Energy barrier as a function of $k_{s}$ for a spherical particle cut from an sc lattice. The particle contains $\mathcal{N}=20479$ spins and has the uniaxial core anisotropy $k_{c}=0.0025$. The solid line is a plot of the analytical expressions (1.1) and (2) from Table 1, using Eqs. (\ref{EcDef}) and (\ref{eq:4thorderEn}).} \label{Bar_sph_sc}
\end{figure}

Fig.~\ref{Bar_sph_sc} shows the energy barrier of a spherical particle cut from an sc lattice as a function of $k_{s}$. The non-monotonic behavior of the energy barrier with $k_{s}$ follows quantitatively that of the EOSP potential (\ref{eq:FullEOSPEn}). Indeed, the solid line in this plot is the analytical results (2) and (1.1) from Table I, using analytical expressions of Eq.~(\ref{eq:4thorderEn}) together with the pure core anisotropy contribution (\ref{EcDef}). The discrepancy at the relatively large $k_{s}$ is due to the fact that the analytical expressions are valid only if the condition (\ref{KcCondition}) is fulfilled; the CSM contribution has not been taken into account.
%
%

\begin{figure}[floatfix]
\includegraphics[width=7.5cm]{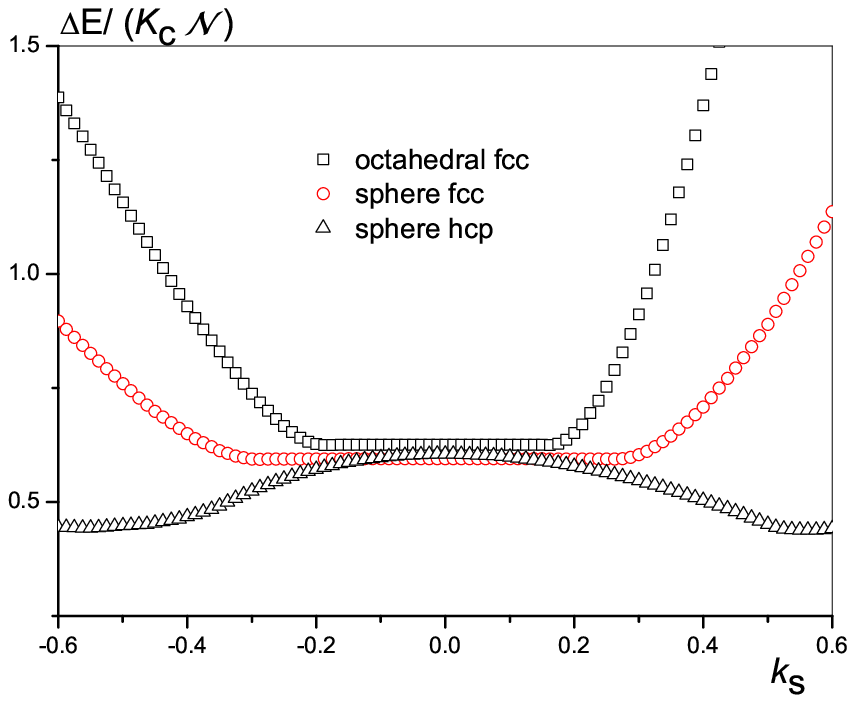}
\caption{Energy barriers against $k_s$ for truncated octahedral particles cut from an fcc lattice ($\mathcal{N}=1688$) and spherical particles cut from the fcc ($\mathcal{N}=1289$) and hcp ($\mathcal{N} = 1261$) lattices. Uniaxial core anisotropy with $k_c=0.0025$ is assumed.}
\label{DiffPart}
\end{figure}

Fig.~\ref{DiffPart} represents the energy barriers against $k_s$ for particles with different shapes and internal structures. First of all, one can see a different dependence on $k_s$ as compared to particles with the sc lattice. In the present case, i.e., $k_\mathrm{ua}^\mathrm{eff}>0, k_\mathrm{ca}^\mathrm{eff}>0$, the energy barriers are given by Eqs.~ (3) and (4.1) in Table I.
%
%
Consequently, for small values of $k_s$ with $|\zeta| < 1$ and neglecting the CSM term, the energy barrier is independent of $k_s$. Accordingly, the nearly constant value of the energy barrier, coinciding with that of the core, is observed for particles in a large range of $k_s$. For larger $k_s$, the energy barrier increases, since $k_\mathrm{ua}^\mathrm{eff}>0$ for particles cut from an fcc lattice. At very large values of $k_s$, i.e., $k_s\gtrsim 100 k_c$ the energy barriers strongly increase with $k_s$ and reach values larger than that inferred from the pure core anisotropy.

\begin{figure}[floatfix]
\includegraphics[width=6.5cm]{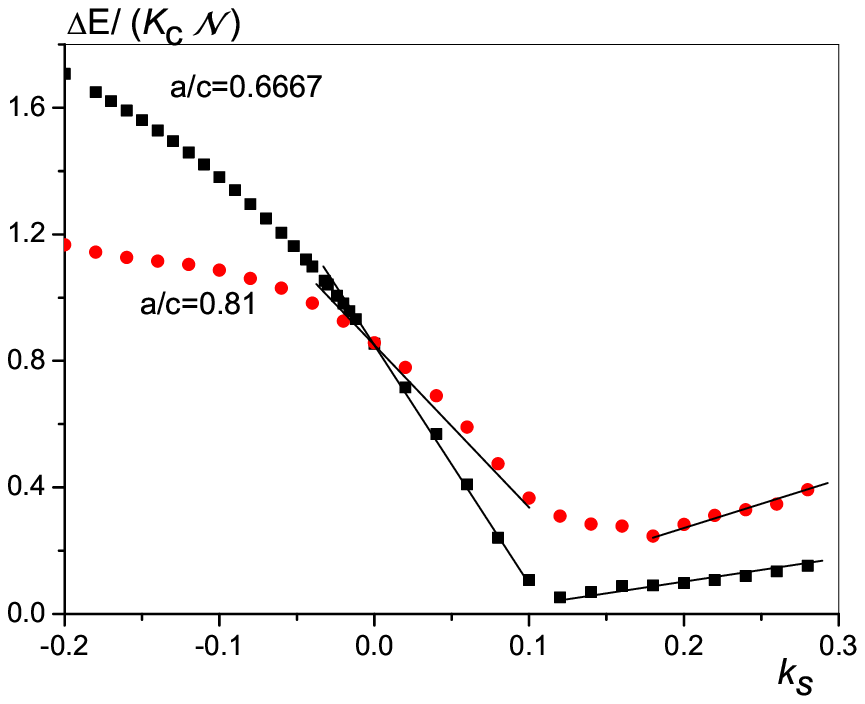}
\caption{Energy barriers versus $k_s$ of ellipsoidal particles with different
aspect ratio ($a/c=0.6667, \mathcal{N}=21121$, and $a/c=0.81, \mathcal{N}%
=21171$, with uniaxial core anisotropy $k_c=0.0025$. The solid lines are
linear fits. }
\label{DeltaE_ellip}
\end{figure}

The energy barriers for ellipsoidal particles are shown in Fig.~\ref{DeltaE_ellip}.
%
%
Note that in this case the effective uniaxial anisotropy constant $k_\mathrm{ua}^\mathrm{eff}$ is a linear function of $k_s$, according to the analytical result (\ref{eq:EnElong}) and the numerical results presented in Fig.~\ref{fig:KMSP_ellipse_fcc_ua_fit}. Here, the energy barriers are not symmetric with respect to the change of sign of $k_s$.
This is due to the fact that for $k_s < 0$ the effective uniaxial constant is a sum of the core anisotropy and the first-order contribution owing to elongation. On the
contrary, when $k_s > 0$, the ``effective core anisotropy" $k_\mathrm{ua}^\mathrm{eff}$ is smaller than the pure core anisotropy $\mathcal{E}_c$ in Eq.~(\ref{EcDef}). This means that at some $k_s$ $k_\mathrm{ua}^\mathrm{eff}$ may change sign. At the same time the effective cubic anisotropy $k_\mathrm{ca}^\mathrm{eff}$ remains positive and is proportional to $k_s^2$. Accordingly, at the vicinity of the point at which $k_\mathrm{ua}^\mathrm{eff}\thickapprox 0$ (see Fig.~\ref{fig:KMSP_ellipse_fcc_ua_fit}), rapid changes of the character of the energy landscape occur. The analysis,
based on the EOSP potential shows that when $k_\mathrm{ua}^\mathrm{eff}>0$ the energy barriers of ellipsoidal particles with sc lattice are defined by Eqs. (2) and (1.1) in Table I and for negative $k_\mathrm{ua}^\mathrm{eff}<0 $ these are given by Eq. (7) and (6.1) in Table I. Note that a regime of linear behavior in $k_s$ exists for both $k_s<0$ and $k_s>0$ (see Fig.~\ref{DeltaE_ellip}), specially when $k_s\lesssim 0.1 (\zeta \ll 1$). In some region of the effective anisotropy constants, e.g., $k_\mathrm{ua}^\mathrm{eff}>0$, $|\zeta|<1$, the energy barrier $\Delta E_\mathrm{EOSP} \approx k_\mathrm{ua}^\mathrm{eff}$, i.e., it is independent of the cubic contribution (ignoring the CSM term). The interval of these parameters is especially large in fcc particles with $k_s<0$, for which $k_\mathrm{ua}^\mathrm{eff}$ does not change sign and the energy barriers are exactly defined by Eq. (3) in Table I.

\subsection{Dependence of the energy barrier on the system size}

As $\mathcal{N} \rightarrow \infty$, the influence of the surface
should become weaker and the energy barriers should recover the
full value $K_c \mathcal{N}$.
\begin{figure}[]
\includegraphics[width=7.5cm]{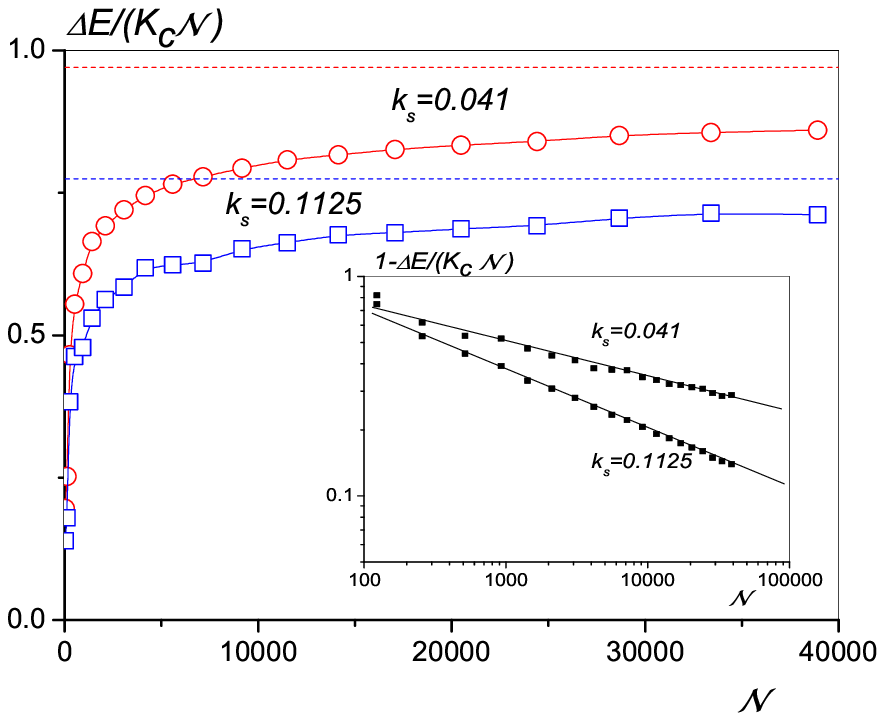}
\caption{Energy barrier as a function of the total number of spins $\mathcal{N}$ for two different values of the surface anisotropy, spherical particles cut from the sc lattice with uniaxial anisotropy $k_c=0.0025$ in the core. The inset shows a slow dependence of the difference between these results and the uniaxial one-particle energy barrier $K_c \mathcal{N}$ in the logarithmic scale. The lines in the inset are the analytical expressions in Eqs.~(\ref{EcDef}) and (\ref{eq:4thorderEn}).} \label{Bar_posKs}
\end{figure}
%
Fig.~\ref{Bar_posKs} shows energy barriers against the total number of spins $\mathcal{N}$ in particles of spherical shape cut from an sc lattice and with two values of $k_s>0$.
First of all, we note that in this case the main contribution to the effective anisotropy consists of two terms: the core anisotropy and the surface second-order contribution (\ref{eq:4thorderEn}). In agreement with this all energy barriers of these particles are always smaller than $K_c\mathcal{N}$ since, as we showed previously for the sc lattice, $k_\mathrm{ca}^\mathrm{eff}$ is negative and the energy barriers in this case are defined by Eq.~(2) in Table I.

Both uniaxial core anisotropy (\ref{EcDef}) and the main
contribution to the effective cubic anisotropy
(\ref{eq:4thorderEn}) scale with $\mathcal{N}$. As
$\mathcal{N}\rightarrow \infty$, the core anisotropy contribution
slowly recovers its full value, i.e.,
$\mathcal{E}_c/(K_c\mathcal{N} )\rightarrow 1$. However, from the
analytical expressions Eqs. (\ref{EcDef}) and
(\ref{eq:4thorderEn}), even when $\mathcal{N} \rightarrow\infty$,
when neglecting the CSM contribution, $\Delta E/(K_c\mathcal{N})$
should approach the value $1 -\kappa k_s^2 /12 k_c$, which is
independent of the system size. Hence, we may conclude that it is
the CSM contribution (\ref {eq:CSM}) that is responsible for the
recovery of the full one-spin uniaxial potential.
Being very small, this contribution produces a very slow increase
of the energy barrier with the system size. In fact, we have
estimated that even spherical particles of diameter $D= 20$ nm (an
estimation based on the atomic distance of $4$\,\AA{}) would have
an effective anisotropy $\Delta E/ (K_c \mathcal{N})$ that is
$13\%$ smaller than that of the bulk.

\begin{figure}[tbp]
\includegraphics[width=7.5cm]{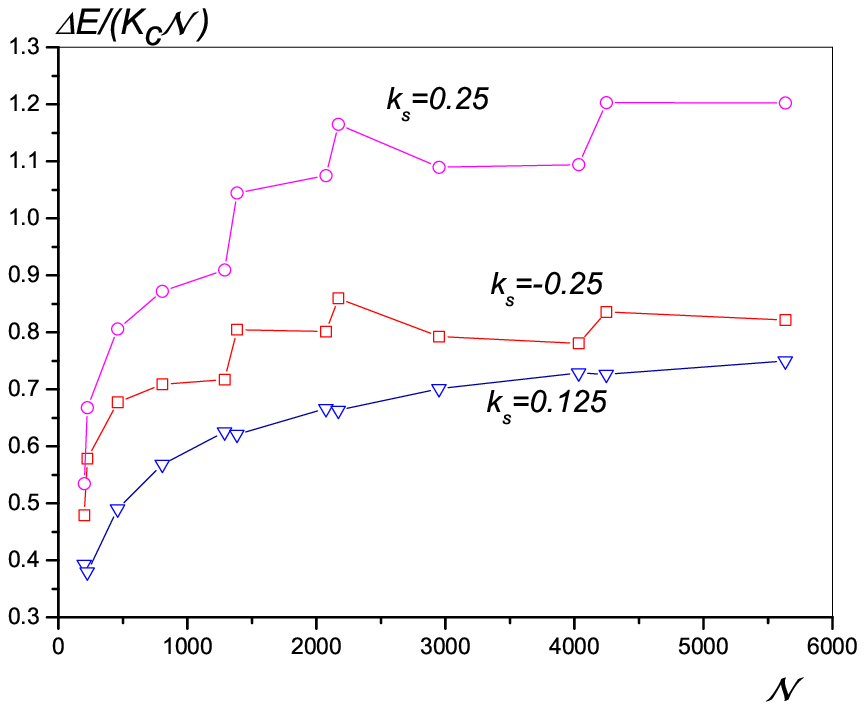}
\caption{Energy barriers versus $\mathcal{N}$ \ for truncated octahedra with
internal fcc structure and uniaxial core anisotropy with $k_{c}=0.0025$.}
\label{Bar_oct_fcc_R}
\end{figure}

Truncated octahedra [see Fig.~\ref{Bar_oct_fcc_R}] show a behavior
similar to that of the spherical particles. The energy barriers in
this case behave very irregularly due to the rough variation of
the number of atoms on the surface. The same effect was observed
in other particles of small sizes.
%
%
For truncated octahedra this effect arises as a consequence of non-monotonic variation of the number of spins on the surface for particles cut from regular lattices. The effective anisotropy of truncated octahedra particles with large $k_{s}>0$ is larger than the core anisotropy in accordance with the fact that $k_{\mathrm{ca}}^{\mathrm{eff}}$ is positive for fcc structures and the energy barriers are defined by Eqs.~(4.1) and (5.1) in Table I.

\begin{figure}[tbp]
\includegraphics[width=7.5cm]{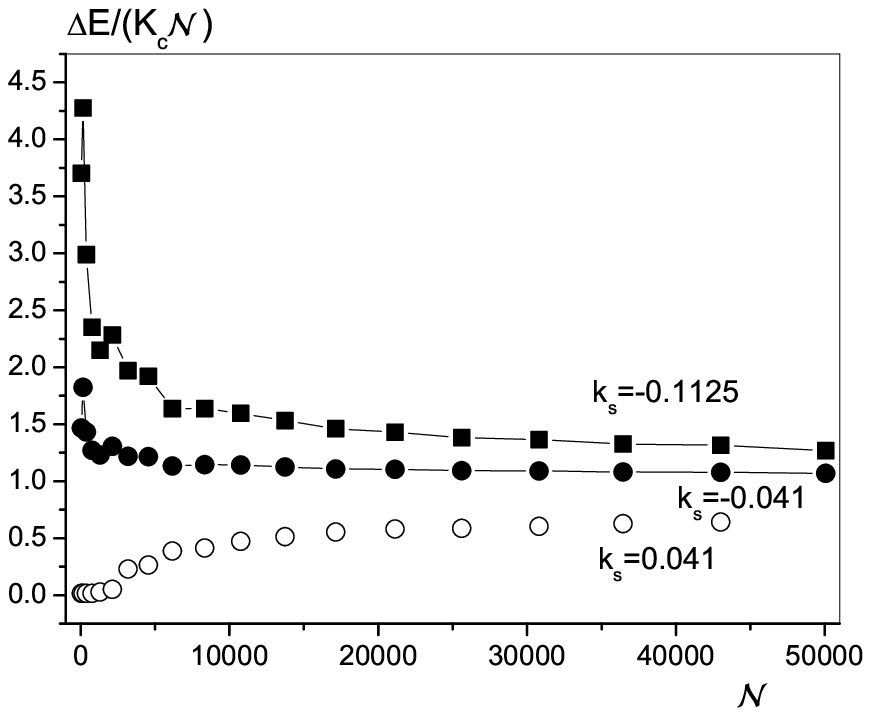}
\caption{Energy barriers as a function of the particle size for ellipsoidal particles with internal sc structure, uniaxial anisotropy $k_{c}=0.0025$ and different values of $k_{s}$.}
\label{Bar_sc_ellips_N}
\end{figure}

Finally, in Fig.~\ref{Bar_sc_ellips_N} we present the energy
barriers of ellipsoidal particles with different values of $k_{s}$
and internal sc structure. The energy barriers in this case are
defined by formulas (1.1) and (2) in Table I. The main
contribution comes from the effective uniaxial anisotropy. The
correction to it due to elongation is positive when $k_{s}<0$ and
negative in the opposite case. Consequently, particles with
$k_{s}<0$ have energy barriers larger than that inferred from the
core anisotropy, and for those with $k_{s}>0$ the energy barriers
are smaller. In this case, the energy barrier approximately scales
with the number of surface spins $N_{s}$ (see Fig.~\ref{Linfit}),
in agreement with the first-order contribution from elongation
(\ref{eq:CSM}).
%
%

\subsection{Applicability of the formula $\mathcal{K}_\mathrm{eff} =
\mathcal{K}_\infty + 6 \mathcal{K}_s/D$}

The results presented above show that in the most general case,
studied here, of a many-spin particle with NSA, this formula is
not applicable, for the following reasons: (i) It assumes that the
overall anisotropy of the particle remains uniaxial. However, we
have shown that the surface anisotropy induces an additional cubic
contribution. (ii) It assumes that the surface anisotropy always
enhances that in the core. In the previous section we saw that
both situations can arise. (iii) It is implicitly based on the
hypothesis that the core and surface anisotropies are additive
contributions. As we have seen above for large $k_s$ (Table I) the
energy barrier indeed can be represented as a sum of the effective
cubic and uniaxial anisotropies. However, the cubic anisotropy
term is proportional to $k_s^2$, which is inconsistent with
formula (\ref{K_eff}). (iv) It assumes a linear dependence of
energy barriers on the parameter $1/D$, or equivalently
$N_s/\mathcal{N}$.

\begin{figure}[]
\includegraphics[width=7.5cm]{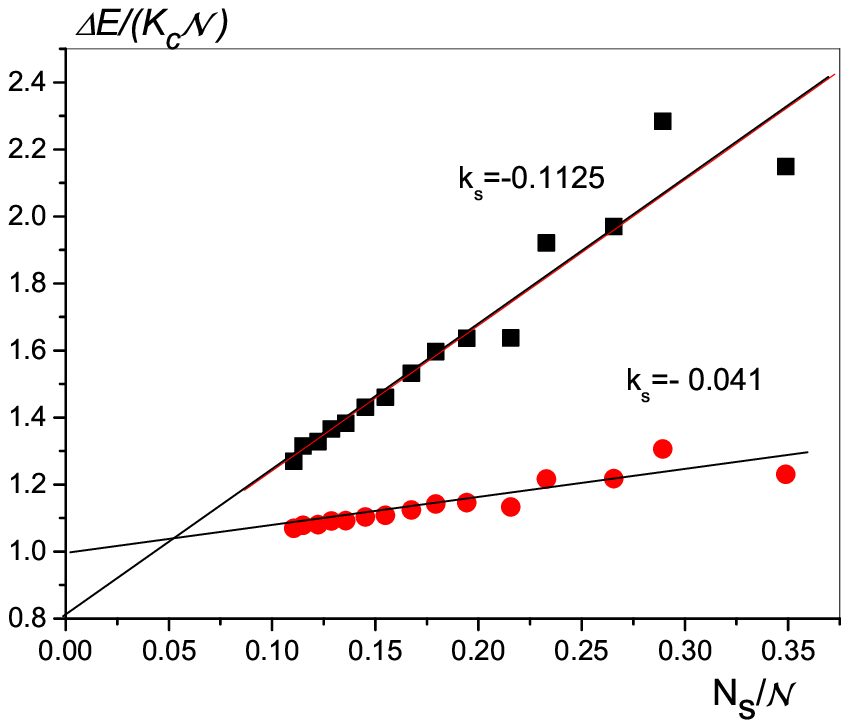}
\caption{Linear fit of energy barriers, versus the surface-to-total number
of spins $N_S/\mathcal{N}$, of ellipsoidal particles with aspect ratio 2:3
with $k_s = -0.041$ (circles) and $k_s = -0.1125$ (squares).}
\label{Linfit}
\end{figure}

Consequently, spherical or octahedral particles cannot be described by
formula (\ref{K_eff}), since in this case (i) No term linear in $k_s$ is
obtained. (ii) No term scales as the ratio of the surface-to-volume number
of spins $N_s/\mathcal{N}$. However, in the case of elongated particles with
a not too large surface anisotropy, (i.e., $|\zeta| < 1$ for fcc lattice or $|\zeta| \ll 1$ for sc lattice), the energy barriers are independent of the
effective cubic anisotropy. In this case, for weakly ellipsoidal particles,
for example, we may write
\begin{equation}  \label{ellipsioidal}
\Delta E_{EOSP} = k_\mathrm{ua}^\mathrm{eff}\approx k_c N_c/\mathcal{N} +
A|k_s|/\mathcal{N}^{1/3}
\end{equation}
where $A$ is a parameter that depends on the particle elongation
and surface arrangement, and which is positive for $k_s<0$ and
negative in the opposite case. Hence, the behavior is as predicted
by formula (\ref{K_eff}). An approximately linear behavior in
$N_s/\mathcal{N}$ was also observed in the case of large surface
anisotropy $\zeta>>1$ (see Fig.~\ref{Bar_OSP}). However, when $\mathcal{N}\rightarrow \infty$, the ``uniaxial anisotropy term" $\mathcal{K}_{\infty}$ is modified by the effective cubic
anisotropy $k_\mathrm{ca}^\mathrm{eff} \thicksim k_s^2$. In
Fig.~\ref {Linfit} we plot the energy barriers of small
ellipsoidal particles with sc
structure, aspect ratio 2:3, and $k_s<0$ from Fig.~\ref{Bar_sc_ellips_N}. %
%
For such particles, formula (\ref{K_eff}) should be modified as $\mathcal{K}_%
\mathrm{eff} = \mathcal{K}{\infty} + |\mathcal{K}_s|N_s/\mathcal{N}$.
Accordingly, in Fig.~\ref{Linfit} we plot the energy barrier against $N_s/%
\mathcal{N}$. These data are highly linear, especially when small
particle sizes are removed, as shown by the fit in
Fig.\ref{Linfit}. We note that in the case of relatively small
surface anisotropy $k_s=-0.041$ (though 17 times larger than in
the core), the full core anisotropy $\mathcal{K}_{\infty} = K_c/v$
( $v$ is the atomic volume) can be extracted. However, for the
larger surface anisotropy $k_s=-0.1125$, $\mathcal{K}_{\infty}$ is
renormalized by the surface contribution (\ref{eq:4thorderEn}). On
the other hand, it is not possible to extract the value of $k_s$,
since the exact proportionality coefficient of Eq.~(\ref{eq:CSM})
is dependent on the particles surface arrangement and elongation.
The effective anisotropy constant $\mathcal{K}_s$ obtained from
this fit is much smaller than the input value, namely, for
$k_s/k_c = 45$ we obtain from the fit
$(k_s/k_c)_\mathrm{eff}=4.3$.

\section{Conclusions}

We have studied in detail effective anisotropies and energy barriers of small magnetic particles within the N\'eel's surface anisotropy model. The present calculations have been performed in a many-spin approach allowing deviations of the spins from the collinear state. They show that the magnetic behavior of small particles is rich and strongly dependent on the particle surface arrangement. The particular structure of each particle and the strength of its surface anisotropy makes each particle unique and its magnetic properties different in each case.

Our calculations show that the magnetic behavior of nanoparticles
with N\'eel surface anisotropy and underlying cubic lattices is
consistent with the effective model of one-spin particle with
uniaxial and cubic anisotropies. The strength of this additional
cubic anisotropy is dependent on many parameters, including the
shape and elongation of the particles, and the underlying crystal
structure which produces a different surface arrangement. The
analytical results have made it possible to classify the various
surface contributions and their effects as: (i) First order
contribution from elongation (\ref{eq:EnElong}), which produces an
additional uniaxial anisotropy, is proportional to $k_s$ and
scales with the number of surface atoms. (ii) Surface second-order
contribution (\ref{eq:4thorderEn}) which is cubic in the net
magnetization components, proportional to $k_s^2$ and scales with
the particle's volume; (iii) core-surface mixing contribution
(\ref{eq:CSM}), which is smaller than the other two contributions
and scales with both surface and volume of the particle.

For particles with sc lattices we compared analytical and numerical calculations for many-spin particles obtaining a very good agreement. Numerical modelling of particles with other structures has confirmed the general character of these effects. The possibility to describe a variety of many-spin particles by a macroscopic magnetic moment with effective uniaxial and cubic anisotropies constant opens a unique possibility to model a collection of
small particles in a multiscale manner, taking into account surface effects through the effective potential (\ref{eq:FullEOSPEn}).

Several very interesting effects were observed in particles with strong surface anisotropy. Particles with magneto-crystalline uniaxial anisotropy develop cubic anisotropy. At the same time the uniaxial anisotropy is also modified by the surface anisotropy. In ellipsoidal particles, surface anisotropy can change the sign of the effective uniaxial anisotropy. Some signatures of these behaviors can be found in the literature. For example in
Ref.~\onlinecite{jametetal01prl}, the magnetic behavior of Co particles with fcc structure and, presumably, cubic magneto-crystalline anisotropy, have demonstrated the effect of uniaxial anisotropy.

The energy barriers of many-spin particles have been evaluated using the Lagrangian-multiplier technique. Their behavior could be well understood with the help of the effective one-spin potential (\ref {eq:FullEOSPEn}). The energy barriers larger than $K_c \mathcal{N}$ have been obtained for all particles with very large surface anisotropy, $k_s \gtrsim 100 k_c$, or for elongated particles with $k_s < 0$. This confirms a well-known fact that the surface anisotropy may contribute to the enhancement of the thermal stability of the particle. However, in the case $k_s>0$, the strength of the surface anisotropy has to be very strong.

We have found that the effective anisotropy extrapolated from the energy barriers measurements is consistent with formula (\ref{K_eff}) only
for elongated particles. In the case of relatively weak surface anisotropy $k_s$, the value of the core anisotropy could be correctly recovered.
However, for larger $k_s$ this effective uniaxial anisotropy $\mathcal{K}_{\infty}$ is renormalized by the surface. The applicability conditions of
formula (\ref{K_eff}) are never fulfilled in spherical or truncated octahedral particles. The control of the parameters governing effective anisotropies does not seem to be possible in real experimental situations and therefore, the extraction of the parameters based on formula (\ref{K_eff}) in some cases may be unreliable.

On the other hand, we should note here that the conclusions of our work have been drawn on the basis of the N\'eel anisotropy model [see also Ref.~\onlinecite{kacbon06prb} for the case of transverse surface anisotropy]. We would like to emphasize that the model itself has been based on microscopic considerations.
Although later there appeared some attempts to justify the model (see e.g. Ref.~\onlinecite{vicmac93prbrc}), in most of cases it remains ad-hoc and not based on the real spin-orbit coupling considerations. A more adequate approach to model magnetic properties of nanoparticles should involve first principle calculations of the magnetic moments and MAE with atomic resolution, like in Co/Cu \cite{xiebla02prb, xieblack04jpcm}.
However at the present state of the art this task remains difficult. Moreover, it is not clear how such models could be used to calculate, for example, thermal properties of small particles. The multiscale hierarchical approach \cite{mryasov05epl} proposes to incorporate the ab-initio calculations into classical spin models. We conclude that more work on theory is necessary to understand surface magnetism, especially in relation with small particles.

\acknowledgments{This work was supported by a joint travel grant
of the Royal Society (UK) and CSIC (Spain) and by the Integrated
Action Project between France (Picaso) and Spain. The work of R.Yanes and
O.Chubykalo-Fesenko has been also supported by the project
NAN-2004-09125-C07-06 from the Spanish Ministry of Science and
Education and by the project NANOMAGNET from Comunidad de Madrid.
D. A. Garanin is a Cottrell Scholar of Research Corporation.}

%

\end{document}